\documentstyle{mn}
\input psfig.sty
\newif\ifAMStwofonts
\newcommand{\figstart}[1]  {\begin{figure} \psfig{#1}}
\newcommand{\lfigstart}[1] {\begin{figure*} \psfig{#1}}
\newcommand{\figend} {\end{figure}}
\newcommand{\lfigend} {\end{figure*}}
\newcommand\lsim{\mathrel{\hbox{\rlap{\hbox{\lower4pt\hbox{$\sim$}}}\hbox{$<$}}}}
\newcommand\gsim{\mathrel{\hbox{\rlap{\hbox{\lower4pt\hbox{$\sim$}}}\hbox{$>$}}}}

\ifoldfss
  \ifCUPmtlplainloaded \else
    \NewTextAlphabet{textbfit} {cmbxti10} {}
    \NewTextAlphabet{textbfss} {cmssbx10} {}
    \NewMathAlphabet{mathbfit} {cmbxti10} {} % for math mode
    \NewMathAlphabet{mathbfss} {cmssbx10} {} %  "   "    "
  \fi
  \ifAMStwofonts
    \ifCUPmtlplainloaded \else
      \NewSymbolFont{upmath} {eurm10}
      \NewSymbolFont{AMSa} {msam10}
      \NewMathSymbol{\upi}     {0}{upmath}{19}
      \NewMathSymbol{\umu}     {0}{upmath}{16}
      \NewMathSymbol{\upartial}{0}{upmath}{40}
      \NewMathSymbol{\leqslant}{3}{AMSa}{36}
      \NewMathSymbol{\geqslant}{3}{AMSa}{3E}

      \let\leq=\leqslant 
       
    \fi
  \fi
\fi % End of OFSS

\ifnfssone
  \newmathalphabet{\mathit}
  \addtoversion{normal}{\mathit}{cmr}{m}{it}
  \addtoversion{bold}{\mathit}{cmr}{bx}{it}
  \newmathalphabet{\mathbfit} % math mode version of \textbfit{..}
  \addtoversion{normal}{\mathbfit}{cmr}{bx}{it}
  \addtoversion{bold}{\mathbfit}{cmr}{bx}{it}
  \newmathalphabet{\mathbfss} % math mode version of \textbfss{..}
  \addtoversion{normal}{\mathbfss}{cmss}{bx}{n}
  \addtoversion{bold}{\mathbfss}{cmss}{bx}{n}
  \ifAMStwofonts
    \ifCUPmtlplainloaded \else
      \UseAMStwoboldmath
      \makeatletter
      \new@mathgroup\upmath@group
      \define@mathgroup\mv@normal\upmath@group{eur}{m}{n}
      \define@mathgroup\mv@bold\upmath@group{eur}{b}{n}
      \edef\UPM{\hexnumber\upmath@group}
      \new@mathgroup\amsa@group
      \define@mathgroup\mv@normal\amsa@group{msa}{m}{n}
      \define@mathgroup\mv@bold\amsa@group{msa}{m}{n}
      \edef\AMSa{\hexnumber\amsa@group}
      \makeatother
      \mathchardef\upi="0\UPM19
      \mathchardef\umu="0\UPM16
      \mathchardef\upartial="0\UPM40
      \mathchardef\leqslant="3\AMSa36
      \mathchardef\geqslant="3\AMSa3E

      \let\leq=\leqslant 

    \fi
  \fi
\fi % End of NFSS release 1

\ifnfsstwo
  \DeclareMathAlphabet{\mathbfit}{OT1}{cmr}{bx}{it}
  \SetMathAlphabet\mathbfit{bold}{OT1}{cmr}{bx}{it}
  \DeclareMathAlphabet{\mathbfss}{OT1}{cmss}{bx}{n}
  \SetMathAlphabet\mathbfss{bold}{OT1}{cmss}{bx}{n}
  \ifAMStwofonts
    \ifCUPmtlplainloaded \else
      \DeclareSymbolFont{UPM}{U}{eur}{m}{n}
      \SetSymbolFont{UPM}{bold}{U}{eur}{b}{n}
      \DeclareSymbolFont{AMSa}{U}{msa}{m}{n}
      \DeclareMathSymbol{\upi}{0}{UPM}{"19}
      \DeclareMathSymbol{\umu}{0}{UPM}{"16}
      \DeclareMathSymbol{\upartial}{0}{UPM}{"40}
      \DeclareMathSymbol{\leqslant}{3}{AMSa}{"36}
      \DeclareMathSymbol{\geqslant}{3}{AMSa}{"3E}

      \let\leq=\leqslant 

    \fi
  \fi
\fi % End of NFSS release 2

\ifCUPmtlplainloaded \else
  \ifAMStwofonts \else % If no AMS fonts
    \def\upi{\pi}
    \def\umu{\mu}
    \def\upartial{\partial}
  \fi
\fi

\title[Inner halo structure I: convergence]
{The Inner Structure of $\Lambda$CDM Halos I: A Numerical Convergence Study} 
\author[C. Power et al.]
{C.~Power,$^1$\thanks{Email: Chris.Power@durham.ac.uk} J.~F.~Navarro,$^{2,5}$ A.~Jenkins,$^1$ C.~S.~Frenk,$^1$ S.~D.~M.~White,$^3$\\  
\newauthor V.~Springel,$^3$ J.~Stadel$^4$ and T.~Quinn.$^4$ \\
$^1$ Dept Physics, University of Durham, South Road, Durham, DH1 3LE \\
$^2$ Dept Physics and Astronomy, University of Victoria, Victoria, BC, V8P 1A1, Canada\\
$^3$ Max-Planck Inst. for Astrophysics, Garching, Munich, D-85740, Germany\\
$^4$ Dept of Astronomy, University of Washington, Seattle, WA 98195, USA.\\
$^5$ CIAR Scholar and Alfred P. Sloan Research Fellow}
\date{submitted to MNRAS}
\pagerange{\pageref{firstpage}--\pageref{lastpage}}
\pubyear{2002}

\begin{document}
\label{firstpage}

\maketitle

\begin{abstract}
We present a comprehensive set of convergence tests which explore the
role of various numerical parameters on the equilibrium structure of a
simulated dark matter halo. We report results obtained with two
independent, state-of-the-art, multi-stepping, parallel N--body codes:
{\tt PKDGRAV} and {\tt GADGET}.  We find that convergent mass profiles
can be obtained for suitable choices of the gravitational softening,
timestep, force accuracy, initial redshift, and particle number. For
softenings chosen so that particle discreteness effects are
negligible, convergence in the circular velocity is obtained at radii
where the following conditions are satisfied: (i) the timestep is much
shorter than the local orbital timescale; (ii) accelerations do not
exceed a characteristic acceleration imprinted by the gravitational
softening; and (iii) enough particles are enclosed so that the
collisional relaxation timescale is longer than the age of the
universe.  Convergence also requires sufficiently high initial
redshift and accurate force computations.  Poor spatial, time, or
force resolution leads generally to systems with artificially low
central density, but may also result in the formation of artificially
dense central cusps.  We have explored several adaptive time-stepping
choices and obtained best results when individual timesteps are chosen
according to the local acceleration and the gravitational softening
($\Delta t_i \propto (\epsilon/a_i)^{1/2}$), although further
experimentation may yield better and more efficient criteria. The most
stringent requirement for convergence is typically that imposed on the
particle number by the collisional relaxation criterion, which implies
that in order to estimate accurate circular velocities at radii where
the density contrast may reach $\sim 10^6$, the region must enclose of
order $3000$ particles (or more than a few times $10^6$ within the
virial radius).  Applying these criteria to a galaxy-sized
$\Lambda$CDM halo, we find that the spherically-averaged density
profile becomes progressively shallower from the virial radius
inwards, reaching a logarithmic slope shallower than $-1.2$ at the
innermost resolved point, $r \sim 0.005 \, r_{200}$, with little
evidence for convergence to a power-law behaviour in the inner
regions.

\end{abstract}
\begin{keywords}
cosmology:theory - dark matter - gravitation
\end{keywords}
\setcounter{footnote}{1}
\section{Introduction}
\label{sec:intro}
Over the past few decades, cosmological N-body simulations have led to 
impressive strides in our understanding of structure formation in universes
dominated by collisionless dark matter. Such simulations have provided an 
ideal test-bed for analytic theories of structure formation, and have been 
used to validate and motivate a variety of theoretical insights into the 
statistics of hierarchical clustering (e.g., Press \& Schechter 1974, 
Bardeen et al. 1986, Bond et al. 1991, Lacey \& Cole 1993, Mo \& White 1996).
 In particular, N-body simulations have played a pivotal role in providing a 
clear framework within which the CDM cosmogony may be compared with 
observation, and in establishing Cold Dark Matter (CDM) as the leading 
theory of structure formation (Davis et al. 1985).

This work has led to the development of a robust theoretical framework which
provides an accurate statistical description of structure growth through 
gravitational instability seeded by Gaussian primordial density fluctuations.
 It is now possible to predict with great accuracy, and based only on the 
initial power spectrum of the primordial fluctuations, a number of important
 statistics that characterize the large scale structure of the universe;
 e.g., the mass function and clustering of dark matter halos and their 
evolution with redshift
(e.g., Jing 1998, Sheth \& Tormen 1999, Jenkins et al. 2001) the non-linear
evolution of the dark matter power spectrum and correlation functions (e.g.,
Hamilton et al. 1991, Peacock \& Dodds 1996), as well as the topological
properties of the large scale structure (e.g., Gott, Weinberg \& Melott 1987).

The impact of such simulation work has been greatest in the non-linear regime,
where analytic calculations offer little guidance.  Recently, and as a result of
the development of efficient algorithms and of the advent of massively parallel
computers, it has been possible to apply N-body studies to the investigation of
structure on small, highly non-linear scales. These studies can now probe scales
comparable to the luminous radii of individual galaxies, thus enabling direct
comparison between theory and observation in regions where luminous dynamical
tracers are abundant and easiest to observe.  Predicting the structure of dark
matter halos on kpc and sub-kpc scales, where it can be compared directly with
observations of galactic dynamics, is one of the premier goals of N-body
experiments, and there has been steady progress in this area over the past few
years.

Building upon the early work of Frenk et al. (1985, 1988), Quinn, Salmon \& Zurek
(1986), Dubinski \& Carlberg (1991) and Crone, Evrard \& Richstone (1993),
Navarro, Frenk \& White (1996, 1997, hereafter NFW) found that, independently of
mass and of the value of the cosmological parameters, the density profiles of
dark matter halos formed in various hierarchical clustering cosmogonies were
strikingly similar. This `universal' structure can be characterized by a
spherically-averaged density profile which differs substantially from the simple
power law, $\rho(r) \propto r^{-\beta}$, predicted by early theoretical studies
(Gunn \& Gott 1972, Fillmore \& Goldreich 1984, Hoffmann \& Shaham 1985, White
\& Zaritsky 1992). The profile steepens monotonically with radius, with
logarithmic slopes shallower than isothermal (i.e. $\beta < 2$) near the centre,
but steeper than isothermal ($\beta>2$) in the outer regions.

NFW proposed a simple formula,
\begin{equation}
\label{eq:nfw}
{\rho(r) \over \rho_{\rm crit}} = {\delta_c \over (r/r_s)(1+r/r_s)^2},
\end{equation}
which describes the density profile of any halo with only two parameters, a
characteristic density contrast{\footnote{We use the term `density contrast'
to denote densities expressed in units of the critical density for closure,
$\rho_{\rm crit}=3H^2/8\pi G$. We express the present value of Hubble's
constant as $H(z=0)=H_0=100\, h$ km s$^{-1}$ Mpc$^{-1}$}}, $\delta_c$, and a scale
radius, $r_s$. Defining the mass of a halo as that contained within $r_{200}$,
the radius of a sphere of mean density contrast $200$, there is a single
adjustable parameter that fully describes the mass profile of halos of given
mass: the `concentration' ratio $c=r_{200}/r_s$.

For the sake of this discussion, the two main points to note from the work of
NFW are the following: (i) the density profile in the inner regions of the halo
is shallower, and in the outer regions steeper, than isothermal, and (ii) there
is no well defined value for the {\it central} density of the dark matter, which
can in principle climb to arbitrarily large values near the centre. 

Conclusion (i) is important, since it is a feature of dark halo models that is
required by observations. For example, it implies that the characteristic speeds
of dynamical tracers may be lower near the centre than in the main body of the
system, as observed in disk galaxies, where the velocity dispersion of the bulge
is lower than indicated by the maximum rotation speed of the surrounding disk,
as well as in galaxy clusters, where the velocity dispersion of stars in the
central cluster galaxy is lower than that of the cluster as a whole. Conclusion
(ii) is also important, since there have been a number of reports in the
literature arguing that the shape of the rotation curves of many disk galaxies
rules out steeply divergent dark matter density profiles (Flores \& Primack
1994, Moore 1994, de Blok et al. 2001, but see van den Bosch \& Swaters 2001), a
result that may signal a genuine crisis for the CDM paradigm on small scales
(see, e.g., Sellwood \& Kosowsky 2000, Moore 2001).

These general results of the work by NFW have been confirmed by a number of
subsequent studies (Cole \& Lacey 1996, Fukushige \& Makino 1997, Huss, Jain \&
Steinmetz 1999, Moore et al. 1998, Jing \& Suto 2000), although there is some
disagreement about the innermost value of the logarithmic slope. Moore et al. 
(1998), Ghigna et al.(2000), and Fukushige \& Makino (1997, 2001) have argued
that density profiles diverge near the centre with logarithmic slopes
considerably steeper than the asymptotic value of $\beta=1$ in NFW's
formula. Kravtsov et al. (1998), on the other hand, initially obtained much
shallower inner slopes ($\beta \sim 0.7$) in their numerical simulations, but
have now revised their conclusions; these authors now argue that CDM halos have
steeply divergent density profiles but, depending on evolutionary details, the
slope of a galaxy-sized halo at the innermost resolved radius may vary between
$-1.0$ and $-1.5$ (Klypin et al. 2001).

Since steep inner slopes are apparently disfavoured by rotation curve data it is
important to establish this result conclusively; if confirmed, it may offer a
way to falsify the CDM paradigm on small scales. Unfortunately, observational
constraints are strongest just where theoretical predictions are least
trustworthy. For example, the alleged disagreement between observed rotation
curves and cuspy dark halo models is most evident for sub-$L_{\star}$ galaxies
on scales of $\sim 1\, h^{-1}$ kpc or less. For typical circular speeds of $\sim
100$ km s$^{-1}$, this corresponds to regions where the density contrast exceeds
$\sim 10^6$. Orbital times in these regions are of order $10^{-3}$ of the age of
the universe, implying that N-body codes must be able to follow particles
accurately for several thousand orbits. Few cosmological codes have been tested
in a systematic way under such circumstances. Furthermore, the cold dark matter
halos that host typical disk galaxies are thought to extend out to a few hundred
kpc, implying that the $\sim$kpc scale probed by observations involves a very
small fraction of the mass and volume of the dark halo. As a consequence, these
regions are vulnerable to numerical artifacts in N-body simulations stemming,
for example, from the gravitational softening or the number of particles.

Extreme care is thus needed to separate numerical artifacts from the true
predictions of the Cold Dark Matter model. In order to validate or `rule out'
the CDM cosmogony one must be certain that model predictions on the relevant
scales are accurate, robust, and free of systematic numerical
uncertainties. Although there have been some recent attempts at unravelling the
role of numerical parameters on the structure of simulated dark matter halos,
notably in the work of Moore et al. (1998), Knebe et al. (2000), Klypin et al.
(2001) and Ghigna et al. (2000), the conclusions from these works are still
preliminary and, in some cases, even contradictory. 

To cite an example, Moore et al. (1998) argue that the smallest resolved scales
correspond to about half the mean inter-particle separation within the virial
radius, and conclude that many thousands of particles are needed to resolve the
inner density profile of dark matter halos. Klypin et al. (2001), on the other
hand, conclude that mass profiles can always be trusted down to the scale of the
innermost $\sim 200$ particles, provided that other numerical parameters are
chosen wisely. Ghigna et al. (2000) suggest an additional convergence criterion
based on the gravitational softening length scale, and argue that convergence is
only achieved on scales that contain many particles and that are larger than
about $\sim 3$ times the scale where pairwise forces become
Newtonian. Understanding the origin of such disparate conclusions and the
precise role of numerical parameters is clearly needed before a firm theoretical
prediction for the structure of CDM halos on $\sim$kpc scales may emerge.

Motivated by this, we have undertaken a large series of numerical simulations
designed to clarify the role of numerical parameters on the structure of
simulated cold dark matter halos. In particular, we would like to answer the
following question: what regions of a simulated dark matter halo in virial
equilibrium can be considered reliably resolved?  This question is particularly
difficult because of the lack of a theory with which the true structure of dark
halos may be predicted analytically, so the best we can do is to establish the
conditions under which the structure of a simulated dark halo is independent of
numerical parameters. This is the question which we endeavor to answer in this
paper.

There is a long list of considerations and numerical parameters that may
influence the structure of simulated dark halos:
\begin{itemize}
\item{the N-body code itself}
\item{the procedure for generating initial conditions}
\item{the accuracy of the force computation}
\item{the integration scheme}
\item{the initial redshift} 
\item{the time-stepping choice}
\item{the gravitational softening}
\item{the particle number}
\end{itemize}
Clearly the list could be substantially longer, but the items above
are widely considered the most important concerning the structure of
simulated dark halos.
%\begin{verbatim}
%\begin{figure}
% \vspace{5.5cm}
% \caption{An example figure in which space has
%          been left for the artwork.}
% \label{sample-figure}
%\end{figure}
%\end{verbatim}
Before we proceed to analyze their role, we must decide which
properties of a dark matter halo we will assess for numerical
convergence. Because, as mentioned above, disk galaxy rotation curves
seem to pose at present one of the most pressing challenges to the CDM
paradigm on small scales, we have decided to concentrate on the
spherically-averaged mass profile, as measured by the radial
dependence of the circular velocity, $V_c(r)=\sqrt{GM(r)/r}$, or,
equivalently, by the inner mean density profile, ${\bar
\rho}(r)=3M(r)/4\pi r^3$. 

We note that the convergence criteria derived here apply strictly only to these
properties, and that others, such as the three-dimensional shape of halos, their
detailed orbital structure, or the mass function of substructure halos, may
require different convergence criteria.

The basic philosophy of our convergence testing procedure is to select
a small sample of halos from a cosmological simulation of a large
periodic box and to resimulate them varying systematically the
parameters listed above, searching for regions in parameter space
where the circular velocity curves are independent of the value of the
numerical parameters, down to the smallest scales where Poisson
uncertainties become important, i.e., roughly down to the radius that
contains $\sim 100$ particles. 

Overall, this is a fairly technical paper of interest mostly to practitioners of
cosmological N-body simulations. Readers less interested in numerical details
may wish to skip to \S~\ref{sec:vprof}, where we discuss in detail the converged
inner mass profile of the galaxy-sized $\Lambda$CDM halo used in our convergence
study. The more technical sections include:
\begin{itemize}
\item
{a discussion of the N-body codes used in this work, initial conditions setup,
and analysis procedure (\S~\ref{sec:method} and Appendix)}
\item
{a general discussion of the consequences of discreteness effects on simulations
of dark matter halos, including a derivation of ``optimal'' choices (for given
particle number) of the timestep and the gravitational softening
(\S~\ref{sec:nepsdt})}
\item
{a comparison between single- and multi-timestepping techniques
(\S~\ref{sec:multistep}) }
\item
{a discussion of the role of the gravitational softening, the initial redshift,
the force accuracy, and the particle number on the inner mass profile of
simulated halos (\S~\ref{sec:numpar}) }
\end{itemize}
Finally, a worked example of how to choose optimal parameters for a
high-resolution simulation is presented in \S~\ref{ssec:workex}. We summarize
our main conclusions in \S~\ref{sec:conclusions}.

\section{Numerical Methods}
\label{sec:method}
\subsection{N-body Codes}
\label{ssec:codes}

Most simulations reported in this paper have been performed with the
parallel N-body code {\tt GADGET}, written by Volker Springel, and
available from {\tt{http:\slash\slash www.mpa-garching.mpg.de\slash gadget}}
(Springel, Yoshida \& White 2001). In order to test the dependence of our
results on the particular algorithmic choices made in {\tt GADGET}, we have also
used {\tt PKDGRAV}, a code written by Joachim Stadel and Thomas Quinn
(Stadel 2001). As we discuss in \S~\ref{sec:nepsdt} and \S~\ref{sec:numpar},
the two codes give approximately the same results for appropriate choices of the
numerical parameters. We have not attempted to carry out a detailed comparison
of the relative efficiency or speed of the codes; such comparison is heavily
dependent on the particular architecture of the hardware used, and on a variety
of optimization and tuning procedures.  We do note, however, that neither code
seems obviously to outperform the other when our strict numerical convergence
criteria are met.

The two N-body codes share a number of similarities. They both evaluate
accelerations (`forces') on individual particles due to all others using a
hierarchical tree data structure (Barnes \& Hut 1986, Jernigan \& Porter 1989),
and (optionally) use individually adaptive time-stepping schemes to advance the
integration of each particle. Periodic boundary conditions are handled in both
codes via Ewald's summation technique (Hernquist, Bouchet
\& Suto 1991), although the implementation of the algorithm in each code is
different.

Gravitational softening is introduced in the form of a `spline' mass
distribution (see, e.g., Hernquist \& Katz 1989, Navarro \& White 1993) which,
unlike the more traditional `Plummer' softening of the early generation of
N-body codes (see, e.g., Aarseth 1985), converges for pairwise interactions
exactly to the Newtonian regime at a finite radius. The length scale of the
spline kernel, $\epsilon_i$, can be chosen individually for each particle in
{\tt PKDGRAV}. {\tt GADGET}, on the other hand, allows for different softenings
to be chosen for up to six different particle `species'. We quote the values of
$\epsilon_i$ so that gravitational interactions between two particles are fully
Newtonian for separations larger than $2\, \epsilon_i$.

The codes differ substantially in their implementation of the tree construction,
in the force-evaluation algorithms and in the integrator scheme. Whereas {\tt
PKDGRAV} uses a spatial binary tree for gravity calculations, {\tt GADGET} uses
a version of the Barnes-Hut geometric oct-tree. Distant tree-node
contributions to the force calculations include up to quadrupole expansion terms
in {\tt GADGET}, but up to hexadecapole in {\tt PKDGRAV}. The tree is rebuilt
every timestep in the version of {\tt PKDGRAV} that we tested (although this is
not the case in the most up-to-date version), whereas we rebuild the tree in
{\tt GADGET} dynamically after $\sim 0.1 \, N_{tot}$ force computations since
the last full reconstruction. ($N_{tot}$ is the total number of particles in the
simulation.)

Finally, {\tt GADGET} uses a simple second-order DKD (drift-kick-drift)
leap-frog integrator scheme with expansion factor as the integration variable,
whereas {\tt PKDGRAV} adopts a cosmic time-based KDK (kick-drift-kick)
algorithm. All integrations are carried out in comoving coordinates.  Details of
these codes may be found in Springel et al. (2001), and in Stadel (2001). In the
following subsections we describe the numerical setup used for the two
codes. All simulations have been run on the IBM-SP supercomputer facilities at
the University of Victoria (Canada), and on the T3Es at the Edinburgh Parallel
Computer Centre (U.K.) and at the Max-Planck Rechenzentrum in Garching
(Germany).

\subsubsection{{\tt GADGET}}
\label{sssec:gadget}

{\tt GADGET} has been the main simulation code used in this study, and it
evolved as the project unfolded from the first public release v1.0 to the latest
available release v1.1. All of the results presented here have been obtained
with the latest version of the code.

{\tt GADGET} presents the user with a number of options regarding time-stepping
choices and the accuracy of the force calculations. In all cases we have used
the tree node-opening criterion recommended by Springel et al. (2001), where a
Barnes-Hut opening criterion with $\theta=0.6$ is used for the first force
computation and a dynamical updating criterion is used subsequently. In this
criterion, a node is opened if $M\, l^4 > f_{\rm acc} \, a_{old} \, r^6$, where
$M$ is the mass of the node, $l$ is the node-side length, and $a_{old}$ is the
acceleration that the particle experienced in the previous timestep.  The
parameter $f_{\rm acc}$ (called {\tt ErrTolForceAcc} in {\tt GADGET}'s parameter
list) is set to $10^{-3}$ in our standard calculations. This condition can be
overridden if the {\tt -DBMAX} compile-time flag is activated. Enabling this flag
imposes an additional condition for node-opening: multipole expansion of a node
is only used if, in addition to the previous condition, the particle is {\it
guaranteed} to lie outside the geometric boundaries of the node in
question{\footnote{A similar condition is activated by default in {\tt
PKDGRAV}}}.  The results reported in \S~\ref{ssec:facc} indicate that these
choices are important to ensure convergence: resolving the inner structure of
dark halos requires highly accurate forces.

{\tt GADGET} uses an integrator with completely flexible timesteps. The code
carries, for each particle, a time, $t_i$, position, ${\bf r}_i$, velocity,
${\bf v}_i$, acceleration, ${\bf a}_i$, gravitational softening, $\epsilon_i$,
and, optionally, a local density, $\rho_i$, and a local one-dimensional velocity
dispersion, $\sigma_i$. From these quantities, timesteps, $\Delta t_i$, can be
computed for each particle according to several possible choices:
\begin{equation}
\label{eq:dti}
\Delta t_i=
\cases{
{\eta_{a\hskip-0.5pt\epsilon}\sqrt{\epsilon_i/a_i}},& if {\tt DtCrit}=0;\cr
{\eta_{a}/a_i},& if {\tt DtCrit}=1;\cr
{\eta_{a\hskip-0.5pt\sigma} (\sigma_i/a_i)},& if {\tt DtCrit}=2;\cr
{\eta_{\rho} {(G \rho_i)}^{-1/2}},& if {\tt DtCrit}=3;\cr
{\eta_{\sigma\hskip-0.5pt\rho}\, {\rm min}[{(G \rho_i)}^{-1/2}, (\sigma_i/a_i)]},&
if {\tt DtCrit}=4,\cr}
\end{equation}
where {\tt DtCrit} refers to the runtime input parameter {\tt ErrTolIntAccuracy}
in {\tt GADGET}, and $\eta_{}$ is a dimensionless constant that controls the
size of the timesteps (except for $\eta_a$, which has dimensions of
velocity){\footnote{For convenience we have defined
$\eta_{a\hskip-0.5pt\epsilon}$ to be directly proportional to the size of the
timestep in all cases. For {\tt DtCrit}=0,
$\eta_{a\hskip-0.5pt\epsilon}^2=2\times{\tt ErrTolIntAccuracy}$}}.  For ease of
reference, we shall refer to the various choices for {\tt DtCrit} using the
following mnemonic shorthand: {\tt EpsAcc} for {\tt DtCrit=0}; {\tt VelAcc} for
{\tt DtCrit=1}; {\tt SgAcc} for {\tt DtCrit=2}; {\tt SqrtRho} for {\tt
DtCrit=3}; and {\tt RhoSgAcc} for {\tt DtCrit=4}, respectively

We report below results obtained with several of these choices. Unless
specified, a maximum timestep was imposed so that all particles took {\it at
least} $200$ timesteps during the whole integration. In practice, this limit
affects a very small fraction of the particles in a typical run: resolving the
inner structure of dark halos requires typically several thousand timesteps.

\subsubsection{{\tt PKDGRAV}}
\label{sssec:pkdgrav}

In the {\tt PKDGRAV} runs reported below we have only explored
variations in two parameters: the time-stepping parameter, $\eta$, and
the gravitational softening, $\epsilon_i$. We note, however, that {\tt
PKDGRAV} is a very flexible code that includes a number of choices for
the integrator scheme and time-stepping, and we have by no means
explored all of its options. {\tt PKDGRAV} was mainly used in this
study to verify that the results obtained with {\tt GADGET} are
independent of the code utilized.

All {\tt PKDGRAV} simulations that used individual timesteps were evolved to
$z=0$ using $50$ system timesteps. The system timestep, $\Delta T$, is the
maximum allowed for any particle. Individual particle timesteps are binned in a
hierarchy so that $\Delta t_i=\Delta T/2^n$, where $n$ was allowed to take any
value in the range ($0$,$20$). This allows particles to take up to $\sim 10^8$
timesteps in a run, which means that in practice no significant restrictions
have been placed on the minimum timestep.

Individual particle timesteps were chosen in {\tt PKDGRAV} runs in a
manner analogous to {\tt GADGET}'s {\tt EpsAcc} criterion, i.e.,
$\Delta t_i \leq \eta \sqrt{\epsilon_i/a_i}$, although quantitatively
accelerations differ because of the choice of integration
variables. The parameter $\eta$ specifies the size of the timesteps
and, consequently, the overall time accuracy of the integration.
%As reported
%below, a value of $\eta \lsim 0.15$ ensures results that are independent of the
%size of the timestep in the test cases we considered.
Finally, the force accuracy in {\tt PKDGRAV} is controlled by $\theta$, a
redshift-dependent opening-node criterion. We have chosen for all runs
$\theta=0.55 \, (z>2)$ and $\theta=0.7$ for $z<2$.

% the main
%goal of our {\tt PKDGRAV} runs has been to verify that the results
%obtained with {\tt GADGET} can be reproduced with an independent code
%using the {\it same} initial conditions. As we report below, there is
%excellent agreement between the two codes when parameters are chosen
%to ensure numerically convergent results.

\subsection{The Initial Conditions}
\label{ssec:ics}

Setting up initial conditions that faithfully represent the cosmogony one wishes
to investigate is a{ \it crucial} step in the simulation process and, despite
the popularity of cosmological N-body simulations, there is surprisingly little
detail in the literature regarding how this is tackled by different groups. The
major references on this topic in the refereed literature are the work of
Efstathiou et al. (1985) and the recent papers by Bertschinger \& Gelb (1991),
Pen (1997), and by Bertschinger (2001; see also {\tt http:\slash\slash
arcturus.mit.edu\slash cosmics} and {\tt http:\slash\slash
arcturus.mit.edu\slash grafic}).

Our particular procedure follows closely that described in Efstathiou et al.
(1985) and is described in detail in the Appendix.  It aims to provide a
particle realization of a Gaussian density field with the chosen primordial
power spectrum, $P(k)$, on scales and at redshifts where linear theory is
applicable.

We adopt the $\Lambda$CDM cosmological model, a low-density universe of flat
geometry whose dynamics is dominated at present by a cosmological constant,
$\Omega_0=0.3$, $\Omega_{\Lambda}=0.7$ and $h=0.65$.  We shall assume that the
initial power spectrum is Harrison-Zel'dovich ($P(k)
\propto k$), modified by an appropriate cosmological transfer
function, $T(k)$. For $\Lambda$CDM simulations we have chosen to use the
analytic representation of the transfer function proposed by Bardeen et al.
(1986) with shape parameter $\Gamma=0.2$.

\begin{figure}
\psfig{figure=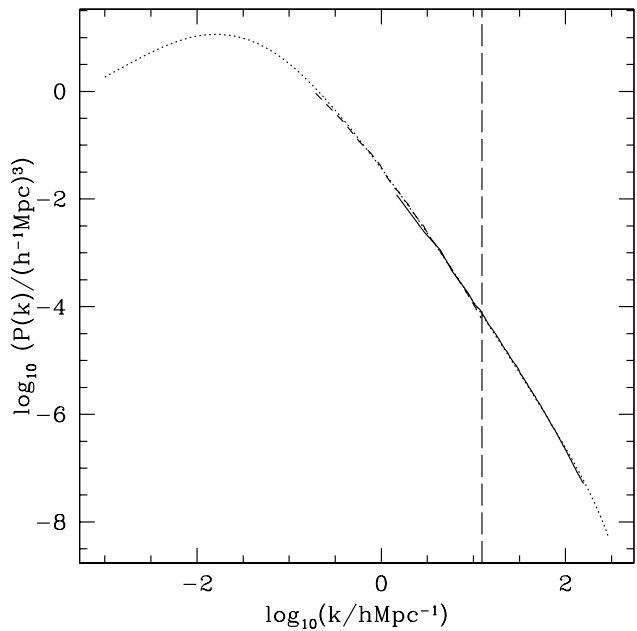,width=250pt,height=250pt}
\caption{ 
The dotted line shows the theoretical $\Lambda$CDM power spectrum at redshift
$z=49$. The short dashed curve shows the measured power spectrum from the
initial conditions of the parent simulation ($L_{\rm box}=32.5 \, h^{-1}$ Mpc,
$N_{\rm box}=128^3$). The solid line shows the power spectrum within the
high-resolution box selected for resimulation ($L_{\rm sbox}=5.08 \, h^{-1}$
Mpc, $N_{\rm sbox}=256^3$). The agreement with the theoretical power spectrum
is good over nearly three orders of magnitude in wavenumber and seven decades in amplitude. Significant departures are expected for both curves at low $k$ as the number of long-wavelength modes is small. The charge assignment scheme
causes a small drop at high-$k$ for both curves.  The vertical long dashed line marks the scale in the resimulated initial conditions which corresponds to the transition between the long waves which are present with the same phase and
amplitude as the parent simulation and the additional short waves added to
improve the resolution.  See Appendix for more details of the computation of
the power spectrum.}
\label{figs:ics}
\end{figure}

Our simulations proceed in two stages. Firstly, a large,
low-resolution, periodic box is run to $z=0$ and used to select halos
targeted for resimulation at much higher resolution (consult the
Appendix for details). For the first step, we have generated a Fourier
representation of the fluctuation distribution on a $128^3$ mesh and
have computed displacements for $128^3$ particles initially arranged
on a cubic grid. The displacements assume an initial redshift of
$z_i=49$ in the $\Lambda$CDM cosmogony and are normalized so that at
$z=0$ the linear rms amplitude of mass fluctuations on spheres of
radius $8\, h^{-1}$ Mpc is $\sigma_8=0.9$. The size of the box is
$L_{\rm box}=32.5 \, h^{-1}$ Mpc (comoving), and the particle mass is
$m_p=4.55 \times 10^9 \, \Omega_0 \, h^{-1} M_{\odot}$.  The dashed
curve in Figure~\ref{figs:ics} shows that the power spectrum computed
from the displaced positions of the $128^3$ particles within this box
is in very good agreement with the theoretical power spectrum (dotted
lines).

The second stage of the initial conditions generating procedure
involves selecting a small region within the large periodic box
destined to collapse into a halo selected for resimulation at higher
resolution. In the case we consider here, this region is a box of
$L_{\rm sbox}=5.08 \, h^{-1}$ Mpc on a side. The advantage of this
procedure is that one can in principle include many more particles in
the high-resolution box than were present in the parent simulation (we
use $N_{\rm sbox}=256^3$ in the case we consider here, giving a
highest-resolution particle mass of $6.5 \times 10^5 \, h^{-1} \,
M_{\odot}$). A new Fourier representation of the theoretical power
spectrum is then generated, retaining the phases and amplitudes of the
Fourier components in the parent simulation and adding waves of higher
frequency, periodic in the high-resolution box, up to the Nyquist
frequency of the high-resolution particle grid.  The solid line in
Figure ~\ref{figs:ics} shows that the power spectrum measured directly
from particle displacements in the high-resolution box is again in
good agreement with the theoretical expectation.

Figure ~\ref{figs:ics} thus demonstrates that the power spectrum is
reproduced well by both the parent simulation and the resimulated
region. Altogether, the power spectrum is fit well over nearly three
decades in wavenumber and seven decades in power.  The maximum
difference between the theoretical power spectrum and the measured
power spectra is less than 0.05 dex, except at low wavenumbers where
the small number of modes makes the variance of the measurement large.

\begin{figure}
\psfig{figure=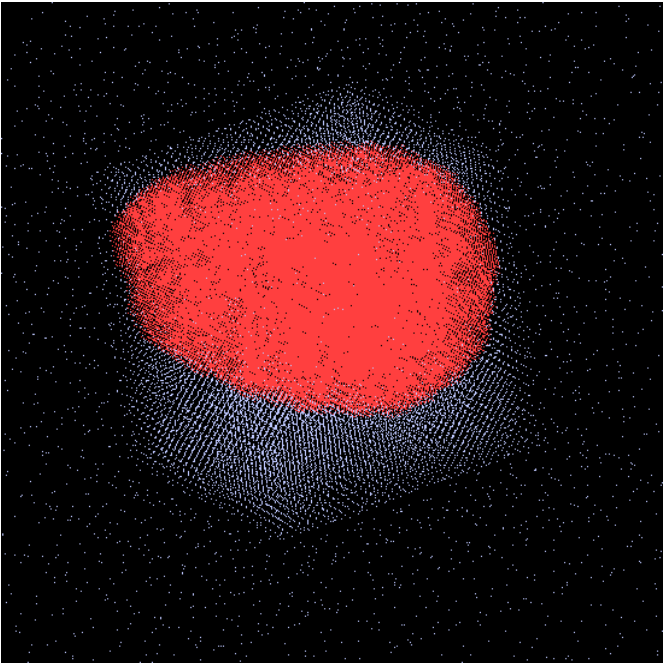,width=250pt,height=250pt}
\caption{Particle distribution in the initial conditions of our $128^3$
runs at $z_i=49$. Clearly seen is the `amoeba'-shaped region containing the
highest resolution particles, which is embedded within the $5.08\,h^{-1}$ Mpc
high-resolution cube used for the resimulation. Beyond the boundaries of the
high-resolution cube lie massive particles that coarsely sample the entire
volume of the $32.5\,h^{-1}$ Mpc periodic box.}
\label{figs:icbox}
\end{figure}

Outside the high-resolution box, we resample the particle distribution in the
parent simulation in order to provide for the tidal forces which act on the high
resolution particles. The resampling procedure bins particles into cells whose
size varies approximately in proportion to their distance from the high
resolution patch, greatly reducing the total number of tidal particles needed to
represent the tidal field. Not all particles in the high-resolution box will end
up near the system of interest, so the location on the original grid of selected
particles is used to identify an `amoeba-shaped' region within the cube that is
retained at full resolution. Regions exterior to the `amoeba' are coarse sampled
with particles of mass increasing with distance from the region of interest
(Figure~\ref{figs:icbox}).

%\vfil 

\subsection{The Simulations}
\label{ssec:sims}

The initial conditions file containing the displacement field for $N_{\rm
sbox}=256^3$ particles generated in the way described above can be easily
rescaled to generate realizations of each system with varying particle number or
starting redshift. To modify the starting redshift, we simply rescale the
displacements and velocities according to the linear growth factor. To reduce
the particle number, we average successively displacements in the
high-resolution box over 8 neighboring cubic cells. We refer to these `reduced'
initial conditions using the total number of particles in the high-resolution
box: $256^3$, $128^3$, $64^3$, and $32^3$, respectively
(Table~\ref{table:numprop}).

These realizations may be used to test how numerical parameters affect
the equilibrium structure of the dark halo at $z=0$. Since runs with
$32^3$ particles are relatively inexpensive, we have used them for a
large series of simulations varying systematically all the numerical
parameters under scrutiny. This series (which contains several hundred
runs) allows us to survey the large available parameter space and to
draw preliminary convergence criteria that are then confirmed with a
series of runs with $64^3$ particles. The $128^3$ and $256^3$
simulations are too expensive to allow a full convergence study, so
fewer of them were carried out, typically using values of the
numerical parameters close to convergence. These are used mainly to
test the dependence of our results on the total number of particles in
the simulations.

\subsection{The Halo}
\label{ssec:halo}

\begin{figure}
\centerline{\psfig{figure=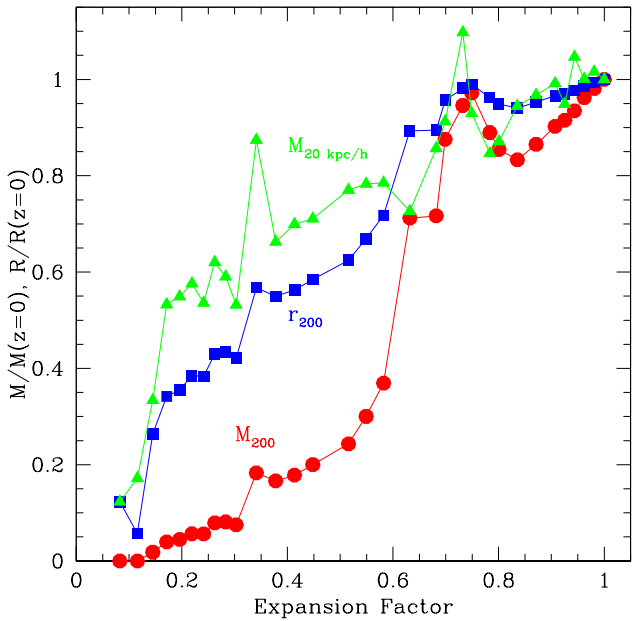,width=250pt,height=250pt}}
\caption{Evolution of the virial radius of the main progenitor of the system, 
$r_{200}$, of the mass contained within that radius, $M_{200}$, and of the 
mass within the innermost $20\, h^{-1}$ (physical) kpc. Data are normalized 
to values at the present day. For the virial radius the ratio of the values 
in {\it comoving} units is shown. The system undergoes its last major merger 
at $z\sim 0.66$, accretes little mass afterwards and is close to virial 
equilibrium at $z=0$. The mass within the inner $20\, h^{-1}$ kpc is assembled
 earlier than the rest, and is only affected seriously during major mergers.}
\label{figs:evmass}
\end{figure}

We concentrate our analysis on a single halo selected from our sample, 
although similar runs on two other halos confirm the conclusions presented 
here. The mass accretion history of this system is presented in 
Figure~\ref{figs:evmass}. The halo accretes half of its present-day mass by 
$z\approx 0.66$ (expansion factor $a=0.6$), when it undergoes a major merger.
 The last significant merger event occurs at $z\approx 0.4$ ($a=0.71$), when 
the system accretes the last $20\%$ of its final mass. After this the system 
remains relatively undisturbed and by $z=0$ it is close to virial equilibrium.
 The virial radius, also shown in Figure~\ref{figs:evmass}, changes by less 
than $7\%$ after $z\approx 0.4$. The mass in the inner regions of the halo is
 assembled much earlier. Half of it is already in place by $z\approx 5$ ($a 
\approx 0.17$) and after $z\sim 1$ substantial fluctuations occur only during
 major mergers. (See the triangles in Figure~\ref{figs:evmass}, which track
 the mass in the innermost $20$ (physical) kpc.)

%\vfil 

\subsection{The Analysis}
\label{ssec:anal}

We focus our analysis on the spherically averaged mass profile at $z=0$. This is measured by sorting particles in distance from the centre and binning them in groups of 100 particles each. The cumulative mass within these bins, $M(r)$, is then used to compute the circular velocity profile of each halo, $V_c(r)=\sqrt{GM(r)/r}$, and the cumulative density profile, ${\bar \rho}(r)=3\, M(r)/4\pi r^3$, which we shall use in our analysis.

It is important to choose carefully the halo centre, especially since the 
halos are not spherically symmetric. The centre of each halo is determined 
using an iterative technique in which the centre of mass of particles within 
a shrinking sphere is computed recursively until a convergence criterion is 
met. At each step of the iteration the centre of the sphere is reset to the 
last computed barycenter and the radius of the sphere is reduced by $2.5\%$.
 The iteration is stopped when a specified number of particles (typically 
either $1000$ particles or $1\%$ of the particles within the high-resolution 
region, whichever is smaller) is reached within the sphere. Halo centers 
identified with this procedure are quite independent of the parameters chosen
 to initiate the iteration, provided that the initial sphere is large enough 
to encompass a large fraction of the system. In a multi-component system, 
such as a dark halo with substructure, this procedure isolates the densest 
region within the largest subcomponent. In more regular systems, the centre 
so obtained is in good agreement with centers obtained by weighing the centre
 of mass by the local density or gravitational potential of each particle. We
 have explicitly checked that none of the results presented here are biased by
 our particular choice of centering procedure.

\section{The relationship between particle number, softening, and timestep}
\label{sec:nepsdt}

The main goal of this study is to identify the conditions under which the
structure of simulated halos, in particular their circular velocity profile, is
independent of numerical parameters. We start with a brief discussion of the
relationship between three of the main parameters: the number of particles, $N$,
the gravitational softening, $\epsilon$, and the timestep, $\Delta t$
(\S~\ref{ssec:anest}). We proceed then (\S~\ref{ssec:sngldt}) to verify
numerically the scalings expected between these quantities through a series of
runs where the timestep for all particles is kept fixed and constant throughout
the evolution.

\subsection{Analytic Estimates}
\label{ssec:anest}

Modeling the formation of dark matter halos with N-body simulations entails a
number of compromises dictated by limited computing resources. The choice of
particle number, timestep, and gravitational softening may all affect, in
principle, the reliability of the structure of simulated halos. We explore here
the various limitations imposed by these numerical parameters.  The analysis
assumes, for simplicity, a steady-state system with circular speed, $V_c(r)$;
enclosed mass, $M(r)=r\, V_c(r)^2/G$; enclosed particle number, $N(r)$; and
orbital timescale, $t_{\rm circ}=2\, \pi \, r/V_c$. The specific energy of a
typical orbit at radius $r$ is $E(r)\approx v^2 \approx V_c(r)^2$.

\subsubsection{$N$ and Collisional Relaxation}
\label{sssec:ncollrellax}

When a finite number of particles is used to represent a system, individual
particle accelerations will inevitably deviate from the mean field value when
particles pass close to each other. Even when orbits are integrated with perfect
accuracy, these `collisions' lead to changes of order unity in energy on the
relaxation timescale (see, e.g., Binney \& Tremaine 1987),
\begin{equation}
\label{eq:trelax0}
{t_{\rm relax} \over t_{\rm circ}} \sim {N(r) \over \ln{(r/\epsilon)}}.
\end{equation}
Thus energy changes due to two-body effects after integration time $t_0$ are
given by
\begin{equation}
{\delta E \over E} \sim \left({t_0 \over t_{\rm relax}}\right)^{1/2} \sim
\left({t_0 \over t_{\rm circ}(r)} {\ln(r/\epsilon) \over N(r)}\right)^{1/2}
\end{equation}
Two-body effects first become important in the inner core of the
system. Suppressing these effects is primarily a condition on the number of
particles and depends only weakly on $\epsilon$. The timestep, of course, does
not appear explicitly in this criterion. We shall return to the limitations
imposed by collisional relaxation in \S~\ref{ssec:npart}.

\subsubsection{Timestep and Integration Accuracy}

Accurate integration of the equations of motion of dark matter particles
requires a careful choice of the timestep adopted to evolve the system.  A
second-order accurate integration with timestep $\Delta t$ induces a relative
error in position, velocity, and energy which scales as
\begin{equation}
\label{eq:errscal1}
{\delta r \over r} \propto {\delta v \over v} \propto {\delta E \over E} \propto \left({v
\Delta t \over r}\right)^3 \propto \left({
\Delta t \over t_{\rm circ}}\right)^3.
\end{equation}
Note that this error depends only on the size of $\Delta t$, and that it is
independent of $N$ and of $\epsilon$, consistent with our assumption of a
smooth, collisionless system.

If errors on subsequent timesteps add incoherently, then the error at the end of
a total integration time, $t_0$, is
\begin{equation}
\label{eq:errscal2}
{\delta E \over E} \propto \left({t_0 \over \Delta t}\right)^{1/2} \left({\Delta t \over
t_{\rm circ}}\right)^3 \propto {(t_0 \Delta t^5)^{1/2} \over t_{\rm circ}^3}.
\end{equation}
For a given $\Delta t$, then, we expect orbits to be reliably modeled only at
radii exceeding a certain value $r_{\rm conv}$ defined by,
\begin{equation}
\label{eq:rconv1}
{t_{\rm circ}(r_{\rm conv}) \over t_0} \propto  \left({\Delta t \over t_0}\right)^{5/6}.
\end{equation}

\subsubsection{Timestep and Discreteness Effects}
\label{sssec:dtrelax}

Finite-$N$ systems are not smooth, and errors in the integration will also 
occur during close encounters between particles. The effects of such 
encounters will be incorrectly treated by the simple integrators used in {\tt
 PKDGRAV} and {\tt GADGET} whenever the predicted separation at mid-step 
between a particle and a near neighbor satisfies $|{\bf s}| = s < v \Delta t$.
 The error in velocity at the end of the step induced by this `unexpected' 
encounter is, then,
\begin{equation}
\label{eq:deltae}
%
%{\delta E \over E} \sim v \Delta t \
%{G \, m \, s/(s^2+\epsilon^2)^{3/2}\over {v^2}},
\delta{\bf v} \sim \Delta t {G m {\bf s}\over (s^2 + 
                       \epsilon^2)^{3/2}}
\end{equation}
assuming Plummer softening. Such encounters occur with probability
\begin{equation}
p(s) \, ds \sim 4 \pi s^2 ds \, {\rho \over m} \sim {s^2\over r^2} \ {ds \over G
m/v^2}.
\end{equation}
where $\rho$ is the mean matter density at the point of encounter, and $m$ is 
the particle mass. The maximum possible size of this error is 
\begin{equation}
%
%\left({\delta E \over E}\right)_{\rm max} \sim {1 \over N(r)} \, {r^2 \over
%\epsilon^2} \, {\Delta t \over t_{\rm circ}}.
\big(\delta v\big)_{\rm max} \sim {G m \Delta t\over \epsilon^2}
\end{equation}
%
%The average energy change obtained by integrating over this distribution is zero
%because the average distribution of the particles corresponds to the mean
%density field of the system. However, the variance in the energy change per step
%is not zero. For a single step,
The average velocity change obtained by integrating eq.~\ref{eq:deltae} over 
the particle distribution is just that due to the mean density field of the 
system. However, averaging the specific energy change over the discrete 
particle distribution gives a positive second-order contribution in excess 
of that expected along the mean-field orbit. For a single step,
\begin{equation}
%
%\delta E^2 \sim {G m v^4 \Delta t^2 \over \epsilon r^2},
\delta E \sim {G m \Delta t^2\over \epsilon r^2}
\end{equation}
where the simplification arises because the integral is dominated strongly by
contributions at $s\sim \epsilon$. After integration time $t_0$ the total 
energy change is then,
\begin{equation}
%
%{\delta E^2 \over E^2} 
{\delta E \over E} 
\sim {t_0 \over \Delta t} \, {G m \Delta t^2 \over
\epsilon r^2} \sim {1\over N(r)} {r\over \epsilon} {t_0 \Delta t \over t_{\rm circ}^2}
\end{equation}
For a given $\Delta t$, then, we expect orbits to be reliably modeled
at radii larger than a certain $r_{\rm conv}$ defined by the following
condition,
\begin{equation}
\label{eq:rconv2}
{t_{\rm circ}(r_{\rm conv}) \over t_0} \approx \left({\Delta t\over
t_0}\right)^{1/2} \, {(Gm/\epsilon)^{1/2} \over V_c(r_{\rm conv}).}
\end{equation}
Since $V_c$ does not change dramatically with radius in CDM halos, we see by
comparing eq.~\ref{eq:rconv1} with eq.~\ref{eq:rconv2} that, in the presence 
of discreteness effects, the number of timesteps required for convergence 
increases as $\epsilon^{-1}$. Economy reasons thus dictate the use of large 
softenings to minimize the number of timesteps. On the other hand, large 
softenings compromise the spatial resolution of the simulations. These 
competing effects suggest the existence of an `optimal' softening choice, 
$\epsilon_{\rm opt}$, which maximizes resolution whilst at the same time 
avoiding discreteness effects and thus minimizing the number of timesteps 
required. We turn our attention to the softening next.

\subsubsection{Softening and Discreteness Effects}
\label{ssec:softdisc}

When accelerations are softened, the {\it maximum} stochastic acceleration 
that can be caused by close approach to an individual particle is roughly
$a_{m\epsilon}=Gm/\epsilon^2$, where $m$ denotes the particle mass. It is 
useful to compare this with the {\it minimum} mean field acceleration, which 
occurs at the outer (virial) radius of the system, $a_{\rm min}\approx 
GM_{200}/r_{200}^2$. The condition $a_{m\epsilon}\lsim a_{\rm min}$ sets a 
lower limit to the softening needed to prevent strong discreteness effects,
\begin{equation}
\label{eq:epsacc}
\epsilon > \epsilon_{\rm acc}\approx {r_{200} \over \sqrt{N_{200}}},
\end{equation}
where $N_{200}=M_{200}/m$ is the total number of particles within
$r_{200}$. When this condition is satisfied, discreteness causes only small
changes in particle accelerations, and so does not significantly affect the
timestepping in integration schemes with an acceleration-based timestep
criterion.

Note that this condition is typically more restrictive than the usual
requirement that large-angle deflections be prevented during two-body
encounters. The latter is given by $\epsilon >
\epsilon_{2b}=Gm/\sigma^2$, where $\sigma$ is the characteristic
velocity dispersion of the system (White 1979). Since $\sigma^2
\approx GM_{200}/2\, r_{200}=G \, m \, N_{200}/2\, r_{200}$, then this
condition requires that forces be softened on scales smaller than $\epsilon_{2b}
\approx 2 \, r_{200}/N_{200}$, which is usually smaller than $\epsilon_{\rm
acc}$. 
%For $\epsilon_{2b}<\epsilon<\epsilon_{\rm acc}$ the stochastic effects of
%discreteness begin to influence first the outer regions, where accelerations are
%low. 
We shall determine the relationship between $\epsilon_{\rm acc}$ and the
`optimal softening' $\epsilon_{\rm opt}$ referred to in \S~\ref{sssec:dtrelax}
empirically in \S~\ref{ssec:sngldt}.

\subsection{Runs with Constant Timestep}
\label{ssec:sngldt}
%\vfil 
\begin{figure}
\centerline{\psfig{figure=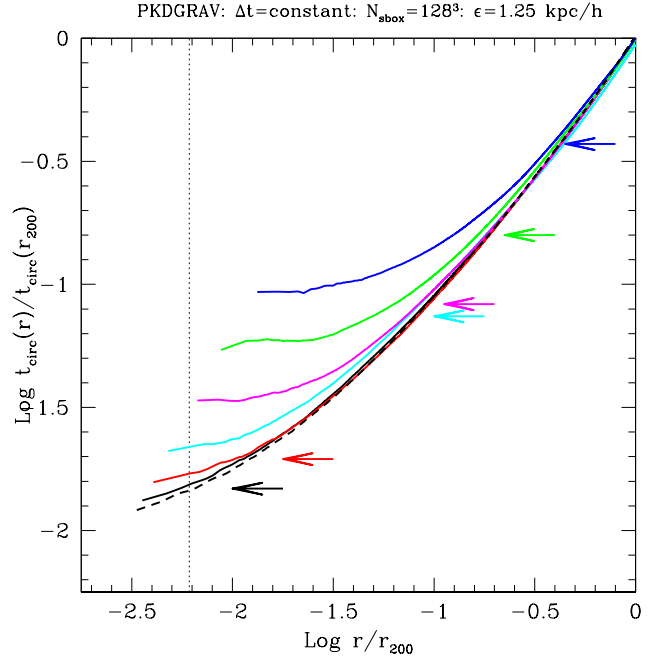,width=250pt,height=250pt}}
\caption{Circular orbit timescale as a function of radius for a series of runs
with constant timestep. All runs have $128^3$ particles within the high-resolution box, $\epsilon=1.25 \, h^{-1}$ kpc (shown with a dotted vertical line), 
and have been run with {\tt PKDGRAV}. The total number of timesteps used in 
each run increases from the top down, from $N_{\Delta t}=100$ to $N_{\Delta t}
=6400$ for the dashed curve at the bottom. From top to bottom, arrows mark the
 smallest radius where convergence, relative to the smallest-timestep run, is
achieved in each case.}
\label{figs:tcprof}
\end{figure}

In order to validate the scalings derived in the previous subsection and to
determine empirically the optimum values of the softening and timestep we have
carried out a series of convergence tests where the timestep has been kept
constant and is shared by all particles. Disabling the multi-timestepping
capabilities of the codes allows us to concentrate on the role of the timestep
size, rather than on the virtues or shortcomings of scaling it in various ways
for individual particles.

The structure of the dark matter halo chosen for our study at $z=0$ is
illustrated in Figure~\ref{figs:tcprof}, where we show the circular orbit
timescale, $t_{\rm circ}(r)=2\pi r/V_c(r)$, as a function of radius.
 Timescales are measured in units of $t_{\rm circ}(r_{200})=0.2 \, \pi \,
 H_0^{-1}$, which is of the order of the age of the universe, $t_0$. Radii
 are measured in units of the virial radius, $r_{200}\approx 205 \, h^{-1}$
 kpc. The gravitational softening, shown by a vertical dotted line, was kept 
constant in these runs,which had $128^3$ high-resolution particles ($\sim 4 
\times 10^5$ within $r_{200}$) and were run with {\tt PKDGRAV}. The innermost
 point plotted in each curve corresponds to the radius that contains $100$
 particles.  From top to bottom, the curves in Figure~\ref{figs:tcprof} 
illustrate how the mass profile of the simulated halo changes as the total
 number of timesteps increases, by successive factors of 2, from 
$N_{\Delta t}=100$ (top curve) to $N_{\Delta t}=6400$ (bottom dashed curve).

%\vfil 
\begin{figure*}
\centerline{\psfig{figure=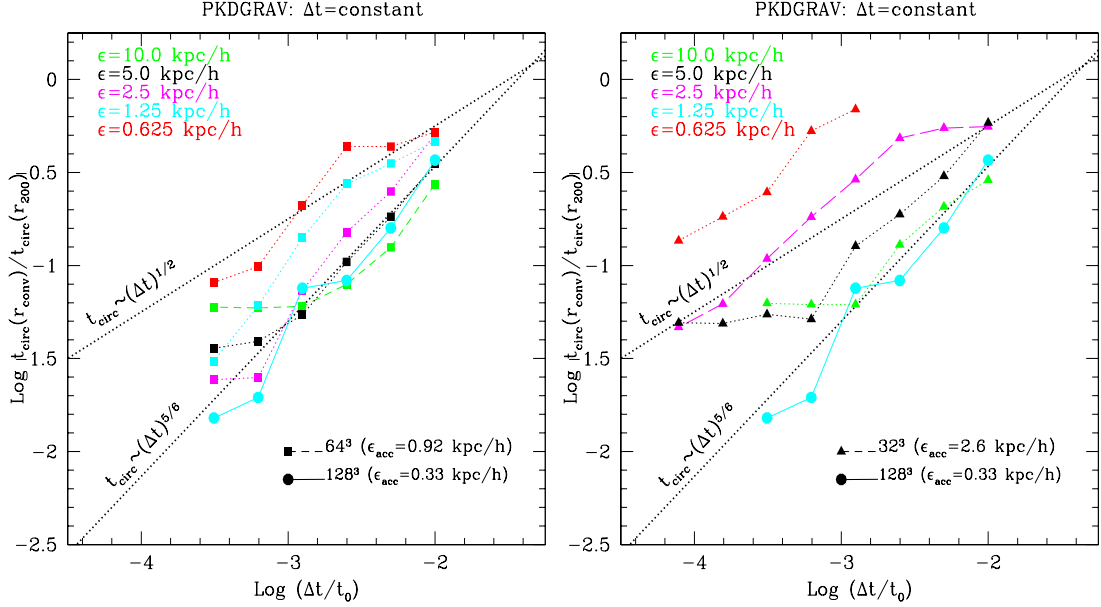,width=417pt,height=250pt}}
\caption{Circular orbit timescale at the smallest `converged' radius (as 
illustrated by arrows in Figure~\ref{figs:tcprof}) as a function of the 
timestep for {\tt PKDGRAV} runs. Left panel shows the results for $N_{\rm sbox
}=64^3$, right panel for $N_{\rm sbox}=32^3$. In both panels the results 
corresponding to $N_{\rm sbox}=128^3$ are shown with filled circles. As the 
timestep decreases, radii where the orbital timescales are shorter become well
 resolved. This `saturates' when the radius becomes comparable to the 
softening and flattens the curves horizontally in some cases. Note also that 
as the softening is reduced the number of timesteps required for convergence 
increases substantially. Refer to text for a thorough discussion.}
\label{figs:epsdt}
\end{figure*}

The halo becomes more centrally concentrated as $N_{\Delta t}$
increases, and approaches a `converged' structure for $N_{\Delta t}
\gsim 3200$. Runs with fewer timesteps than this still converge to the
right mass profile but at increasingly larger radii. It is interesting
to explore how the radius where convergence is achieved, $r_{\rm
conv}$, depends on the number of timesteps. We find $r_{\rm conv}$ by
identifying the radius at which systematic departures greater than
$10\%$ in the circular timescale profile first become noticeable,
gauged against the run with the largest number of timesteps. This is
easily read off the profiles presented in
Figure~\ref{figs:tcprof}. Arrows in this figure indicate $t_{\rm
circ}(r_{\rm conv})$ for each choice of $N_{\Delta t}$.

Filled circles in Figure~\ref{figs:epsdt} show the converged timescales thus
determined as a function of the size of the timestep. Converged circular times
follow closely the $\Delta t^{5/6}$ dependence expected from eq.~\ref{eq:rconv1}, suggesting that the choice of softening in this series is such that 
discreteness effects are negligible. This is perhaps not surprising, as 
$\epsilon=1.25$ kpc/h is about 4 times larger than the lower limit estimated 
in eq.~\ref{eq:epsacc}, $\epsilon_{\rm acc}=r_{200}/\sqrt{N_{200}}=0.32$ kpc/h
 (for $N_{\rm sbox}=128^3$).

Solid squares in Figure~\ref{figs:epsdt} (left panel) correspond to
the same exercise carried out for several choices of the softening
when the number of high-resolution particles is reduced to $64^3$. For
the same softening, $\epsilon=1.25$ kpc/h, achieving convergence with
$64^3$ particles requires significantly smaller timesteps than with
$128^3$, as expected since discreteness effects become more important as
the number of particles is reduced. It is also clear from
Figure~\ref{figs:epsdt} that the dependence of $t_{\rm circ}(r_{\rm
conv})$ on $\Delta t$ changes as $\epsilon$ decreases, shifting
gradually from $\Delta t^{5/6}$ to $\Delta t^{1/2}$. This transition
is precisely what is expected from the analytic estimates in
\S~\ref{ssec:anest} (see eqs.~\ref{eq:rconv1} and \ref{eq:rconv2}).

There is further supporting evidence in Figure~\ref{figs:epsdt} for
the validity of the analytic estimates. Consider for example the right
panel, where we present the results of runs with $32^3$
high-resolution particles. The trends are similar to those in the left
panel, but the transition to the discreteness-dominated regime
($t_{\rm circ}(r_{\rm conv}) \propto \Delta t^{1/2}$) occurs for even
larger values of $\epsilon$.

It is possible to use these results to estimate the softening above which
discreteness effects become unimportant for the various series. From
Figure~\ref{figs:epsdt}, we find that, for $64^3$ particles, this `optimal'
softening is somewhere between $2.5$ and $5$ kpc/h, while for $32^3$ particles
it is of order $\sim 10$ kpc/h. Our $128^3$ runs suggest that $\epsilon_{\rm
opt}\approx 1.25 \, h^{-1}$ kpc for this series. The optimal softening appears
thus to scale with $N$ just as suggested by our discussion of
eq.~\ref{eq:epsacc}. The simple empirical rule,
\begin{equation}
\label{eq:epsopt}
\epsilon_{\rm opt} \approx  4 \, \epsilon_{\rm acc}={4 \, r_{200} \over \sqrt{N_{200}}},
\end{equation}
appears to describe the numerical results well.

The reason why $\epsilon_{\rm opt}$ is about a factor of $4$ larger
than $\epsilon_{\rm acc}$ is likely related to the fact that, when
softening is chosen to optimize results for halos at $z=0$, the
choice is not optimal for their progenitors at earlier times. Indeed,
$r_{200}(M,z) \propto (\Omega(z)/\Omega_0)^{1/3} M^{1/3} (1+z)^{-1}$, which implies that, for
softenings fixed in comoving coordinates, $\epsilon/\epsilon_{\rm
opt}(N,z) \propto N(z)^{1/6}$. Small-$N$ progenitors thus have smaller
softenings than optimal and may be subject to discreteness
effects. The dependence on the number of particles is weak, however,
and it is possible that the factor of $4$ in eq.~\ref{eq:epsopt} may
act as a `safety factor' to ensure that discreteness effects are
negligible at all times.

A number of other predictions from the analytic scalings presented in
\S~\ref{ssec:anest} are also confirmed by the data in
Figure~\ref{figs:epsdt}. For example, when discreteness effects dominate,
converged timescales are expected to scale as $\epsilon^{-1/2}$
(eq.~\ref{eq:rconv2}). This is in good agreement with the results of the
$32^3$-particle runs; for given timestep, $t_{\rm circ}(r_{\rm conv})$ is seen
to increase by roughly a factor of $2$ when $\epsilon$ decreases by a factor of
$4$, from $\epsilon=2.5$ to $0.625$ kpc/h.

Finally, the analytic estimates suggest that the timestep choice should be
independent of $N$ and $\epsilon$ when discreteness effects are
unimportant. This is also reproduced in the simulation series: for
$\epsilon\gsim \epsilon_{\rm opt}$, all runs, independent of $N$, lie along the
{\it same} dotted line that delineates the
\begin{equation}
t_{\rm circ}(r_{\rm conv}) \approx 15 \left({\Delta t\over t_0}\right)^{5/6}
t_{\rm circ}(r_{200})
\label{eq:tcrconv}
\end{equation}
scaling. This confirms that the size of the timestep is the most
important variable when discreteness effects are unimportant; roughly
$400$, $7000$, and $110000$ timesteps are needed to resolve regions
where $t_{\rm circ}$ is, respectively, $\approx 10\%$, $1\%$ and
$0.1\%$ of the orbital timescale at the virial radius.

\subsection{Convergence and integrator schemes}
\label{ssec:convint}

So far these conclusions are based on runs carried out with {\tt PKDGRAV}. Are
 they general or do they depend on the particular choice of integrator scheme?
 We have explored this by performing a similar series of constant-timestep 
runs with {\tt GADGET}, which uses a different integrator (\S~\ref{ssec:codes}
).  There is another difference between the {\tt GADGET} series and the one 
carried out with {\tt PKDGRAV}: {\tt GADGET} integrates the equations of 
motion using the expansion factor, $a$, as the time variable. 
Constant-timestep runs carried out with {\tt GADGET} were therefore evolved 
using a fixed expansion-factor step, $\Delta a$. Comparing {\tt GADGET} and
 {\tt PKDGRAV} runs with the same {\it total} number of steps, {\tt GADGET}
 takes shorter time steps than {\tt PKDGRAV} at high-redshift, longer ones at
 moderate $z$ and similar ones at $z \approx 0$.

\begin{figure}
\centerline{\psfig{figure=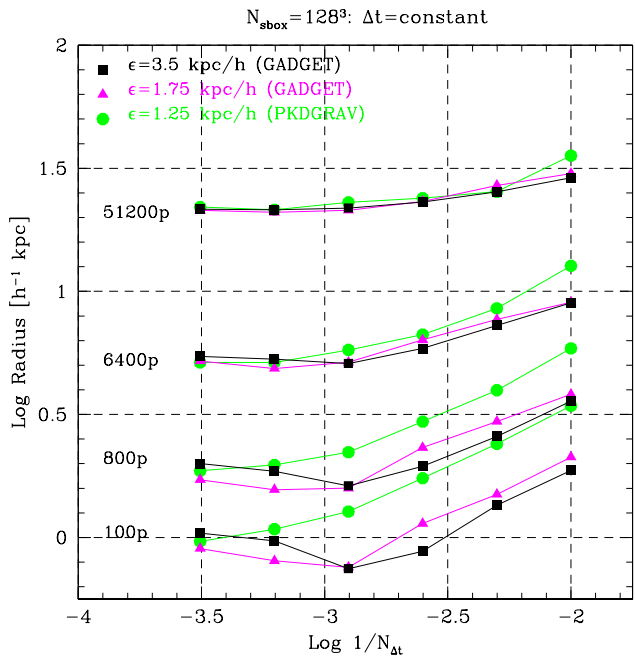,width=250pt,height=250pt}}
\caption{Radii enclosing various numbers of particles as a function of the 
total number of timesteps. Results shown correspond to runs with $128^3$ 
particles in the high-resolution box and were run with {\tt PKDGRAV} and {\tt
 GADGET}, as labeled. {\tt PKDGRAV} runs approach convergence progressively 
and monotonically. On the other hand with poor time resolution {\tt GADGET} 
can produce artificially dense `cuspy' cores, most noticeable when $N_{\Delta
 t}=800$. These `cuspy cores' seem to be inherent to the integrator and 
stepping schemes chosen in {\tt GADGET}. Note, however, that both codes need 
approximately the same number of timesteps for full convergence.}
\label{figs:rsngldt_128}
\end{figure}

We compare the results of the two series in Figure~\ref{figs:rsngldt_128}, 
where we plot, at $z=0$, the radii containing various mass fractions of the 
halo as a function of the number of timesteps, $N_{\Delta t}$. The three 
series shown correspond to runs with $128^3$ high-resolution particles; two 
were run with {\tt GADGET} and one with {\tt PKDGRAV}. The choice of softening
 in each case is indicated in the figure labels.  The four radii shown 
contain, from bottom to top, $0.025\%$, $0.2\%$, $1.6\%$, and $12.8\%$ of the 
mass within $r_{200}$, respectively.

Convergence is approached gradually and monotonically in {\tt PKDGRAV} runs 
(solid circles in Figures~\ref{figs:epsdt} and \ref{figs:rsngldt_128}). For 
$N_{\Delta t} \sim 3200$ convergence is achieved at all radii containing more 
than $\sim 100$ particles; fewer timesteps are needed to converge at larger 
radii, as discussed in the previous subsection.  

%\vfil 

Convergence also occurs gradually, but {\it not} monotonically, in the case of
{\tt GADGET} (solid squares and triangles in Figure~\ref{figs:rsngldt_128}).
For the same number of steps, {\tt GADGET} results typically in mass profiles
that, near the centre, are more concentrated than {\tt PKDGRAV}'s, as shown by
the systematically smaller radii that contain the same mass fraction. The effect
is particularly noticeable for $N_{\Delta t}\approx 800$, when the central
density profile is actually {\it steeper} than the `converged' result achieved
for $N_{\Delta t} \gsim 3200$.

Further runs with different softenings and numbers of particles suggest that the
presence of these `cuspy cores' in systems evolved with poor time resolution is
inherent to $\Delta a=$constant {\tt GADGET} runs, and not just a fluke. On the
other hand, the artificial cusps only occur in regions well inside the
convergence criterion derived from the {\tt PKDGRAV} series. Outside the
convergence radius delineated by eq.~\ref{eq:tcrconv} both {\tt GADGET} and {\tt
PKDGRAV} results appear safe: one may conclude that {\tt GADGET} and {\tt
PKDGRAV} require approximately the same number of timesteps to resolve the whole
system.

To summarize, the central densities of systems evolved with poor time resolution
may be over- or under-estimated, depending on the integrator scheme adopted.
Such sensitivity to the integrator scheme emphasizes the vulnerability of the
central regions to numerical artifact and the need for detailed convergence
studies such as the one presented here before firm conclusions can be reached
regarding the inner density profiles of CDM halos.

\subsection{Summary}

The agreement presented above between numerical results and analytic
estimates gives us confidence that it is possible to achieve
convergence in the mass profiles of simulated dark halos down to
scales which contain as few as $100$ particles or where the
gravitational softening starts to dominate. A few prescriptions for an
efficient and accurate integration seem clear:
\begin{itemize}
\item
choose gravitational softenings so that $\epsilon \gsim \epsilon_{\rm opt}=4\,
r_{200}/\sqrt{N_{200}}$ (eq.~\ref{eq:epsopt}) to minimize the number of
timesteps needed, and
\item
regard as converged only regions where circular orbit timescales exceed $\approx
15 \left({\Delta t/ t_0}\right)^{5/6} t_{\rm circ}(r_{200})$
(eq.~\ref{eq:tcrconv}).
\end{itemize}

One problem with these prescriptions is that, in a large cosmological N-body
simulation, where systems of different mass and size form simultaneously, it is
possible to choose optimal values of the numerical parameters only for systems
of roughly the same mass. Also, resolving the inner density profiles, where
orbital timescales can reach a small fraction of the age of the universe, may
prove impractical with a constant timestep, as the number of timesteps is then
dictated by the densest region of the system, which may contain only a small
fraction of the total number of particles. It is therefore important to learn
how the structure of simulated dark halos is affected when non-optimal choices
of numerical parameters are made as well as when multi-timestepping integration
techniques are adopted.  We turn our attention to these topics in the following
sections.

\section{Adaptive Multi-stepping techniques}
\label{sec:multistep}

In order to improve efficiency, many cosmological N-body codes use individual
timesteps that can vary with time and from particle to particle. This allows the
time integration scheme to adapt spatially so as to achieve high accuracy across
the whole body of non-linear structures. The two codes used in this study, {\tt
PKDGRAV} and {\tt GADGET}, can use individual timesteps, although, as discussed
in \S~\ref{ssec:codes}, they differ significantly in the choice of integration
scheme.

Evaluating the efficiency gain is not straightforward, since computing resources
in most parallel environments do not scale in simple ways with the total number
of particles and of timesteps, and the latter is ill-defined when individual
adaptive timesteps are adopted. We shall assume, for simplicity, that the bulk
of the computational work is invested in computing individual accelerations
(`forces'), and shall deem efficient timestepping choices that achieve `full
convergence'{\footnote{We use the term `full convergence' when it extends down
to the scale containing as few as 100 particles or the gravitational softening,
whichever is larger.}} whilst minimizing the total number of force computations,
$N_{\rm ftot}$.

For the integrators used in {\tt PKDGRAV} and {\tt GADGET} forces are computed
once every time the position (or velocity) of a particle is advanced, so that
${\bar N}_{\Delta t}=N_{\rm ftot}/N$ can be thought of as the average number of
timesteps in a run. $N_{\rm ftot}$ is an imperfect measure of the total
computational work, since it neglects the overhead that stems from tree
construction, neighbor searching (if required by the timestepping choice),
synchronization, and communication between nodes, but is nonetheless a useful
guide for assessing the efficiency of various timestepping techniques.

\subsection{Comparison of timestep criteria}
\label{ssec:compdt}

\begin{figure}
\centerline{\psfig{figure=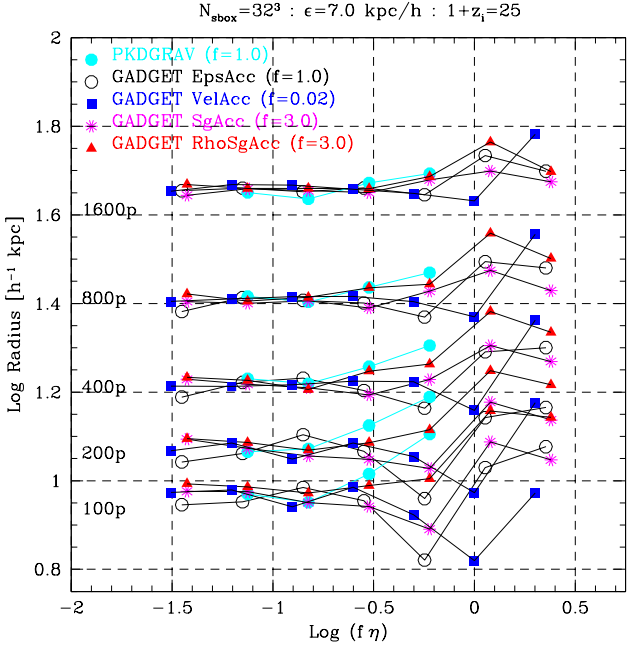,width=250pt,height=250pt}}
\caption{ As Figure~\ref{figs:rsngldt_128}, but for radii as a function of the
time-stepping parameter, $\eta$. The number of timesteps decreases linearly 
with $\eta$ (\S~\ref{eq:dti}). The values of $\eta$ shown in the Figure have 
been scaled by an arbitrary factor, $f$, so that for given $f\, \eta$ all runs
 have similar number of total force computations, $N_{\rm ftot}$. Values of 
$f$ are given in the figure labels.}
\label{figs:rvseta}
\end{figure}

{\tt GADGET} allows for five different ways of setting the timestep, and we 
have explored extensively four of them. Our main results are illustrated in 
Figure~\ref{figs:rvseta}, which is analogous to Figure~\ref{figs:rsngldt_128}
 but for runs with $32^3$ high-resolution particles.  The radii shown enclose
 $1.6\%$, $3.2\%$, $6.5\%$, $12.9\%$, and $25.8\%$ of the mass within the 
virial radius, respectively, and are shown as a function of the timestep 
parameter, $\eta$ (\S~\ref{sssec:gadget}). We adopt for this series a 
softening of $7\, h^{-1}$ kpc, close to the `optimal' value for this number 
of particles (see Table~\ref{table:numprop}). For convenience, we have scaled
 $\eta$ by an arbitrary factor $f$ (listed in the labels of Figure~\ref{figs:rvseta}) chosen so that, for given $f\, \eta$, all runs in this figure incur 
approximately the same total number of force computations. CPU consumption is
 lowest for {\tt EpsAcc} and {\tt VelAcc}, $\sim 25\%$ higher for {\tt SgAcc},
 and highest (by $\sim 60\%$) for {\tt RhoSgAcc} because the neighbor search 
required by the latter two criteria imposes a significant overhead.

\begin{table*}
\begin{center}
\caption{Properties of the simulated halo.}
\begin{tabular}{cccccccc}
Label & $r_{200}$ & $V_{200}$ & $M_{200}$ & $N_{\rm sbox}$ & $N_{200}$ &
$\epsilon_{\rm acc}$ &$\epsilon_{\rm opt}$ \\ & $[h^{-1}$ kpc] & [km s$^{-1}]$ &
$[10^{10}$ M$_{\odot}]$ &&& $[h^{-1}$ kpc] & $[h^{-1}$ kpc] \\
\hline
Halo 1 &205 &205 &200 &$256^3$ & $3.17\times 10^6$& 0.12 & 0.46\\
       &    &    &    &$128^3$ & $3.97\times 10^5$& 0.33 & 1.30\\
       &    &    &    &$64^3$  & $4.96\times 10^4$& 0.92 & 3.68\\
       &    &    &    &$32^3$  & $6.20\times 10^3$& 2.60 & 10.4\\
%Halo 2 &230 &285 &230 \\
%Halo 3 &210 &220 &210 \\
\label{table:numprop}
\end{tabular}
\end{center}
\end{table*}

%\vfil 

The main conclusion to be drawn from Figure~\ref{figs:rvseta} is that all
timestepping choices appear to converge for approximately the same value of $f\,
\eta \lsim 0.2$ or, equivalently, for the same $N_{\rm ftot}$. For $f\eta\approx
0.2$, $N_{\rm ftot} \approx 2.2\times 10^7$, which implies that on average a
minimum of $\sim 650$ timesteps is required for full convergence. This is
comparable to the number of {\it constant} timesteps needed for full convergence
(see Figure~\ref{figs:epsdt}).

For $f\, \eta>0.2$, deviations from convergence are obvious in all
cases. Deviations are monotonic in the case of {\tt PKDGRAV} and {\tt RhoSgAcc}
runs; densities at all radii increase gradually as the timestep decreases and
converge for $f\, \eta \lsim 0.2$. On the other hand, the behaviour of the inner
mass profile in the case of other criteria is clearly non-monotonic: the central
shells dip well below the converged value before bouncing back to convergence as
$f\, \eta$ approaches $0.2$.  This is reminiscent of the artificially cuspy
cores discussed in \S~\ref{ssec:convint}, but it seems to affect radii well
beyond the softening.  

Note that these artificial `cuspy cores' affect runs with {\tt GADGET}'s {\tt
EpsAcc} criterion as well, which is formally the same as used in {\tt
PKDGRAV}. The monotonic approach to convergence seen in {\tt PKDGRAV} runs thus
suggest that the presence of `cuspy cores' in runs with poor time resolution is
an artifact related to {\tt GADGET}'s integrator scheme rather than to the
timestepping choice.  Artificially cuspy cores are an undesirable feature in
large cosmological simulations, because dense cores may survive the hierarchical
assembly of structure and lead to artifacts in the density profiles of systems
formed by the merger of affected progenitors. This kind of subtle artifact again
demonstrates that careful convergence studies of the kind presented here are
needed to guarantee that the inner mass profiles of dark matter halos can be
robustly measured in N-body simulations.

\subsection{The Dependence on Softening}
\label{ssec:softdep}

According to the analysis presented in \S~\ref{ssec:anest}, the
timestep required for convergence is independent of the softening when
discreteness effects are unimportant (i.e., when $\epsilon \gsim
\epsilon_{\rm opt}$, see eq.~\ref{eq:rconv1}) but should become
increasingly short as $\epsilon$ decreases below the optimal value (see
eq.~\ref{eq:rconv2}). Since optimal softenings can only be adopted for systems
of roughly the same mass in a large cosmological simulation, optimizing the
choice for massive clumps leads to less-than-optimal softenings in low-mass
halos. For such systems, keeping the values of $\eta$ found to give convergence
in the last subsection (i.e. $f\, \eta\approx 0.2$ with the values of $f$ given
in Figure~\ref{figs:rvseta}) may not guarantee convergence unless the
timestepping criterion scales appropriately with softening. For fixed $\eta$,
timesteps decrease as $\epsilon^{1/2}$ in PKDGRAV and for the EpsAcc criterion
of GADGET, but are unchanged as the softening decreases for the other GADGET
criteria.

\begin{figure}
\centerline{\psfig{figure=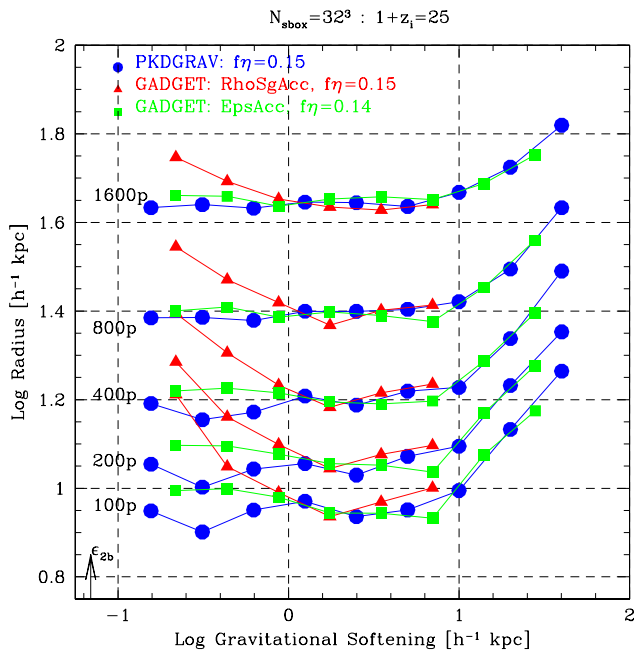,width=250pt,height=250pt}}
\caption{Radii enclosing various mass fractions measured at $z=0$ in our 
$32^3$ simulations as a function of the gravitational softening scale length,
 $\epsilon$. Pairwise interactions become Newtonian at distances exceeding 
$2 \, \epsilon$. The virial radius of the halo is $r_{200}=205\, h^{-1}$ kpc
 and the total number of particles within this radius is $N(r_{200})\approx
 6200$.  Note that for {\tt PKDGRAV} and {\tt EpsAcc} runs the mass profile is
 independent of softening for $\epsilon < 6\, h^{-1}$ kpc, provided that the
 softening remains larger than $\epsilon_{2b}$, the minimum needed to prevent
 large-angle deflections during particle collisions (\S~\ref{ssec:anest}). For
 {\tt RhoSgAcc}, the choice $f\,\eta=0.15$ leads to convergence for $\epsilon 
> 1\, h^{-1}$ kpc, but results in large deviations for smaller softenings. See
 \S~\ref{ssec:softdep} for details.}
\label{figs:rvssoft}
\end{figure}

The effects of this are illustrated in Figure~\ref{figs:rvssoft}, which shows
the result of adopting $f \, \eta \sim 0.2$ whilst gradually reducing the
softening to values almost two orders of magnitude below optimal. For {\tt
RhoSgAcc}{\footnote{For simplicity, we discuss here only {\tt RhoSgAcc}; similar results apply to {\tt VelAcc}.}} (solid triangles), $f\, \eta=0.15$ seems
appropriate for $\epsilon$ close to or slightly smaller than $\epsilon_{\rm
opt}\approx 10 \, h^{-1}$ kpc, but an artificially low density core clearly
develops for softenings well below the optimal value. This behaviour is not seen in the case of {\tt EpsAcc}, where convergence appears firm even for values of $\epsilon$ approaching the large angle-deflection limit, $\epsilon_{2b}$.

%\vfil 

We emphasize that this does {\it not} signal a failure of the {\tt RhoSgAcc}
criterion; rather, it implies that the timesteps chosen with $f\,\eta\sim0.2$
are not short enough to achieve convergence when $\epsilon \ll
\epsilon_{\rm opt}$. 
%(Note that because timesteps in {\tt EpsAcc} depend
%explicitly on softening, the total number of force calculations incurred by this
%criterion when $\epsilon \ll \epsilon_{\rm opt}$ is much larger than by {\tt
%RhoSgAcc} for given $f\,\eta$.) 
Indeed, choosing {\tt RhoSgAcc} and $f\,\eta\approx 0.2
(\epsilon/\epsilon_{opt}) ^{1/2}$ for small softenings eliminates the
artificially low density core shown in Figure~\ref{figs:rvssoft} at a cost in
total number of timesteps, $N_{\rm ftot}/N$, not very different from that
required by {\tt EpsAcc}. This demonstrates clearly the need to take smaller
timesteps when $\epsilon \ll \epsilon_{\rm opt}$.

How should timesteps scale with $\epsilon$? Eq.~\ref{eq:rconv2} suggests a
linear dependence when discreteness effects dominate, $\Delta t \propto
\epsilon$, although the firm convergence seen for {\tt EpsAcc} in
Figure~\ref{figs:rvssoft} indicate that a gentler dependence, $\Delta t \propto
\epsilon^{1/2}$, may actually suffice. This is because the actual individual
timesteps in this criterion are determined by the ratio, $(\epsilon/a_i)^{1/2}$,
and accelerations are high during close encounters when softenings are small. As
a result, the `effective' size of {\tt EpsAcc} timesteps scales roughly linearly
with softening when $\epsilon \ll \epsilon_{\rm opt}$. We have verified this by
comparing the `maximum' number of timesteps, defined by the total number of
timesteps taken by a hypothetical particle which, at all times, has the minimum
timestep of all particles in the system, with the minimum number of {\it
constant} timesteps required for convergence (see \S~\ref{ssec:sngldt}). The
agreement is quite good.

\subsection{Adaptive versus Constant Timestep}
\label{ssec:aconstdt}

Finally, we investigate the computational gain/loss associated with adopting a
constant or adaptive time stepping technique when a criterion such as {\tt
EpsAcc} is selected. Again, we shall assume that the bulk of the computational
work is invested in computing individual accelerations, although this measure
neglects the cost of tree construction. Ordinarily, tree-making contributes a
small fraction of the CPU budget, but this is not necessarily the case in
multiple timestepping schemes when a full tree structure is recalculated every
time particles in the smallest time bin are advanced. This is the case in the
version of {\tt PKDGRAV} that we tested. {\tt GADGET}, on the other hand,
recomputes trees only after a certain number of interactions have been computed
(\S~\ref{ssec:codes}), so the comparison is not straightforward.

We have chosen for the comparison maximally-converged {\tt PKDGRAV} runs, i.e.,
those requiring the minimum number of timesteps for full convergence.  The main
conclusion may be gleaned from Table~\ref{table:ndt}, where we list the total
number of force computations, $N_{\rm ftot}$, for runs with $32^3$
high-resolution particles{\footnote{For ease of comparison, we have not reduced
in this series the number of high-resolution particles through the `amoeba'
procedure described in the Appendix for runs listed in this
Table~\ref{table:ndt}.}}  and three different choices for the gravitational
softening; $\epsilon=10 \, h^{-1}$ kpc ($\approx \epsilon_{\rm opt}$), as well
as $\epsilon=2.5$ and $0.625 \, h^{-1}$ kpc. The number of constant timesteps
needed for full convergence depends sensitively on softening, as discussed in
\S~\ref{sec:nepsdt}; $N_{\Delta t}$ climbs from $800$ to $25600$ as $\epsilon$
decreases from $10$ to $0.625 \, h^{-1}$ kpc. The total number of force
calculations is directly proportional to $N_{\Delta t}$, and increases from $2.6
\times 10^7$ to $8.4 \times 10^8$.

\begin{table*}
\begin{center}
\caption{Properties of maximally-converged runs ({\tt PKDGRAV}).}
\begin{tabular}{ccccccc}
$N_{\rm sbox}$&$\epsilon$&$N_{\Delta t}$&$N_{\rm ftot}$&${\bar N}_{\Delta t}$&$N_{\rm ftot}$\\
                    & $[h^{-1}$ kpc]& (constant)& (constant)& (multiple)   & (multiple) \\
\hline
%$32^3$  & 10.0 & 800 & $2.6 \times 10^7$ & 2541 & $2.1\times 10^7$\\
$32^3$  & 10.0 & 800 & $2.6 \times 10^7$ & 640 & $2.1\times 10^7$\\
%         & 5.0  & 1600& $5.2 \times 10^7$ &      & $??\times 10^7$\\
         & 2.5  & 1600& $5.2 \times 10^7$ & 1342 & $4.4\times 10^7$\\
%         & 1.25 & 3200& $1.0 \times 10^8$ &      & $??\times 10^7$\\
         & 0.625&25600& $8.4 \times 10^8$ & 2777& $9.1\times 10^7$\\
$64^3$ & 2.5 & 3200 & $8.4 \times 10^8$ & 1754 & $4.6 \times 10^8$\\
%         & 1.25 & 3200 & $8.4 \times 10^8$ & 18543 & $6.6 \times 10^8$\\
$128^3$  & 1.25 & 3200 & $6.7 \times 10^9$ & 2956 & $6.2 \times 10^9$\\
\label{table:ndt}
\end{tabular}
\end{center}
\end{table*}

Table~\ref{table:ndt} shows that, when adaptive multiple timesteps are allowed,
the total number of force calculations needed is comparable when $\epsilon \sim
\epsilon_{\rm opt}$, but far fewer when the softening is well below the
`optimal' value. This demonstrates that the small timesteps required when the
softening is well below the `optimal' value are only needed briefly by a small
subset of particles undergoing close encounters. Adaptive multi-stepping schemes
vastly outperform the fixed timestep approach when $\epsilon \ll \epsilon_{\rm
opt}$.

\subsection{Summary}

To summarize, we find that all timestepping criteria we have considered can
deliver convergence at comparable cost. However, the {\tt EpsAcc} criterion is
the one that suffers least from overheads related to computing values for
individual timesteps, and thus appears to be the most efficient of the criteria
explored in this study. We emphasize, however, that this choice is primarily
empirical; further investigation may very well lead to better and more efficient
alternatives than any of the ones considered here.

Further, for softenings close to the `optimal' value, the computational gain
that results from adopting multi-stepping schemes is rather modest, especially
considering that the implementation of multi-stepping incurs a non-negligible
cost in terms of memory usage and bookkeeping. 
%Results for other numbers of
%particles, also listed in Table~\ref{table:ndt}, lend support to this
%interpretation. 
Smaller softenings increase the importance of discreteness effects and lead to
integrations with very small timesteps dictated by occasional
encounters. Multi-stepping schemes are strongly favored under these
circumstances.

\section{The Role of other Numerical Parameters}
\label{sec:numpar}

Proper convergence requires, of course, that appropriate choices be made for
{\it all} relevant parameters. We now turn to the analysis of the separate role
of other numerical parameters. Unless explicitly stated, we will undertake the
analysis of each parameter using only runs for which all other parameters take
`converged' values. This can only be done after a large parameter space search
since the effects of combinations of some parameters may be subtle. For example,
a timestep that is adequate for some gravitational softening may be inadequate
when the softening is substantially modified.  Because of this restriction, the
results in the following subsections contain, for clarity, only a small fraction
of all runs performed.

\subsection{The Gravitational Softening}
\label{ssec:softening}

Large cosmological simulations generally use a single particle mass and thus
resolve systems of different mass with different numbers of particles. This
implies that it is possible to choose `optimal' values of the softening only for
a small range of halo masses, since $\epsilon_{\rm opt} \propto
r_{200}/N_{200}^{1/2} \propto N_{200}^{-1/6}$. This may not be too restrictive
for the resimulations we discuss here, since they focus on one system at a time,
but it does affect significantly large cosmological simulations.  If, for
example, an optimal softening choice is made for the most massive system
expected to form at, say, $z=0$ , it will be smaller than the optimal value for
less massive systems present at the same time (see eq.~\ref{eq:epsopt}). How are
their mass profiles affected and what regions in such systems may be considered
converged?

To address this question, we have undertaken a large series of simulations where
the softening, $\epsilon$, was varied systematically while choosing
`converged' values of all other parameters. We have explicitly checked that,
for example, doubling or halving the number of timesteps (or the initial
redshift) has no appreciable effect and that, for given number of particles, the
results discussed in this subsection depend only on $\epsilon$.

We show the results of this series in Figure~\ref{figs:rvssoft}, where
radii enclosing various mass fractions are shown as a function of
$\epsilon$ in simulations with $32^3$ high-resolution particles. Since
$r_{200} \approx 205 \, h^{-1}$ kpc, the radii shown in
Figure~\ref{figs:rvssoft} probe a large fraction of the halo's radial
extent, between $4\%$ and $22\%$ of the virial radius. For this
system, $\epsilon_{2b}=0.066 \,h^{-1}$ kpc $\approx 3.2\times
10^{-4}\, r_{200}$ and $\epsilon_{\rm opt}\sim 10 \, h^{-1}$ kpc
$\approx 4.9\times 10^{-2} \, r_{200}$.

As Figure~\ref{figs:rvssoft} shows, the mass profiles obtained with
the two codes agree to better than $20\%$ (i.e., to better than $10\%$
in circular velocity) even for radii containing as few as $100$
particles. Full convergence is achieved for a wide range of softening
scales, provided that $\epsilon_{2b}< \epsilon \lsim 6 h^{-1}$
kpc. The mass profiles are essentially unchanged even as the softening
is varied by almost two orders of magnitude.

A second important point to note in Figure~\ref{figs:rvssoft} is that
for $\epsilon \sim 12 \, h^{-1}$ kpc (only slightly larger than
$\epsilon_{\rm opt}$) the profile deviates from the converged one even
as far out as $60 \, h^{-1}$ kpc; i.e., more than $5$ times the
softening length. This contrasts with the results for $\epsilon \sim 6
\, h^{-1}$ kpc, where the mass profile appears to have converged down
to almost one softening length scale.  Clearly, assuming that mass
profiles are affected out to a certain multiple of the softening
length is an oversimplification that is {\it not} supported by these
results.

\begin{figure}
\centerline{\psfig{figure=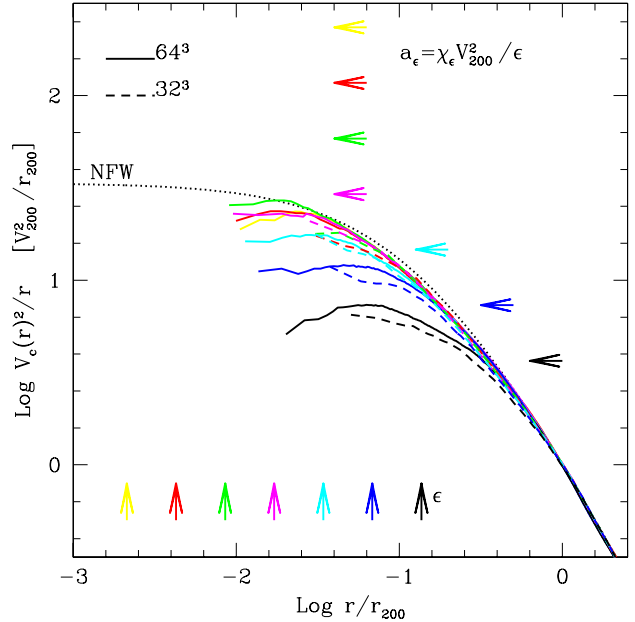,width=250pt,height=250pt}}
\caption{Spherically-averaged `acceleration' profiles ($V_c^2(r)/r$)
for $64^3$ and $32^3$ runs, shown for several choices of the softening
scalelength, $\epsilon$. The dotted line corresponds to the
acceleration profile of an NFW model with concentration $c=10$. The
vertical arrows denote the value of the softening parameter,
$\epsilon$, for each run. The profiles line up, from bottom to top, in
order of decreasing $\epsilon$. As $\epsilon$ approaches $\sim 0.01 \,
r_{200}$, the acceleration profiles converge to a solution similar to
the fiducial NFW curve. Profiles significantly affected by the
softening deviate from the converged result at a radius where the
acceleration matches the characteristic acceleration associated with
the circular velocity of the halo, $V_{200}$, and $\epsilon$:
$a_{\epsilon}=\chi_{\epsilon}V_{200}^2/\epsilon$, with
$\chi_{\epsilon} \approx 0.5$. The values of $a_{\epsilon}$
corresponding to each adopted value of $\epsilon$ are shown by the
horizontal arrows.}
\label{figs:accprof}
\end{figure}

What determines the smallest converged radius for a given softening
length scale? Since softenings introduce a characteristic acceleration
on small scales, it is instructive to consider the mean acceleration
that particles experience as a function of the distance from the
centre of the system. This radial acceleration profile,
$a(r)=GM(r)/r^2=V_c^2(r)/r$, is shown in Figure~\ref{figs:accprof} for
two series of runs where the gravitational softening has been varied
systematically by two orders of magnitude. The values of the softening
in each run are shown with small vertical arrows near the bottom of
the figure. Solid and dashed curves correspond to runs with $64^3$ and
$32^3$ particles in the high-resolution box, respectively. As the
softening is decreased from $\epsilon \sim 0.1\, r_{200}$ by successive
factors of two, the acceleration profiles become steeper and converge
to a unique profile for $\epsilon \lsim 0.03 \, r_{200} \approx 6 \,
h^{-1}$ kpc, as shown in Figure~\ref{figs:accprof}. The convergent
profile is well approximated by an NFW profile with $c=10$, shown by
a dotted line in Figure~\ref{figs:accprof}.

We note two interesting features of the acceleration profiles shown in
Figure~\ref{figs:accprof}. The first is that the effects of softening on the
acceleration profile depend rather weakly on the number of particles used; for
given $\epsilon$, the profiles corresponding to runs with $32^3$ and $64^3$
particles agree reasonably well, and they approach the same `converged' profile
for $\epsilon \lsim 0.03 \, r_{200}$. The second feature is that acceleration
profiles deviate from the `converged' profile near the centre for larger
values of the softening. Interestingly, deviations occur at radii where the
acceleration exceeds a `characteristic' acceleration,
\begin{equation}
a_{\epsilon}=\chi_{\epsilon} V_{200}^2/\epsilon,
\label{eq:aeps}
\end {equation}
which depends only on the circular velocity of the halo and on the
value of the softening adopted. This characteristic acceleration is
shown (for $\chi_{\epsilon}\approx 0.5$) with horizontal arrows in
Figure~\ref{figs:accprof}. The mass profile of a simulated halo
becomes unreliable for accelerations exceeding $a_{\epsilon}$.

This result suggests an empirical interpretation of the effects of softening on
the mass profile of a simulated halo: the choice of gravitational softening
imposes an effective limit on the accelerations that may be adequately
reproduced in the system. This is interesting, since for systems with density
profiles similar to that proposed by NFW, there is a {\it maximum} acceleration
that particles may experience. Indeed, $a(r)=V_c^2/r$ tends to a well-defined
maximum,
\begin{equation}
\label{eq:amax}
a_{\rm max}= {c^2/2 \over \ln(1+c)-c/(1+c)} \,\, {V_{200}^2 \over r_{200}}
\end{equation}
as $r$ approaches zero. If $\epsilon$ is such that 
\begin{equation}
a_{\epsilon}\gsim a_{\rm max},
\label{eq:aepsamax}
\end{equation}
then it appears to impose no substantial restriction on the mass profile. For
example, Figure~\ref{figs:rsngldt_128} shows that the converged mass within
$\sim 1 h^{-1}$ kpc appears not to change as $\epsilon$ varies between $1.25$
and $3.5 h^{-1}$ kpc. At face value, this would appear to imply that the mass
profile can be trusted down to almost one third of the softening length scale
when the condition expressed in eq.~\ref{eq:aepsamax} is satisfied. In order to
be conservative, however, we shall hereafter assume that converged radii cannot
be less than $\epsilon$.

How does the upper limit on $\epsilon$ dictated by this constraint compare 
with $\epsilon_{\rm opt}$, the minimum needed to prevent discreteness effects
 and minimize the number of timesteps? The answer depends on the number of 
particles, as well as on the concentration of the system, and imposes an 
effective lower limit on the number of particles needed to satisfy both 
conditions simultaneously, $N_{200}\gsim (2c)^4/(\ln(1+c)-c/(1+c))^2$. For 
$c\approx 10$, we find that roughly $70,000$ particles within the virial 
radius are needed to carry out a simulation where the softening is small 
enough not to restrict significantly the resolution of the inner mass profile
 and large enough to prevent discreteness effects from hindering the 
computational efficiency of the calculation.

%This result has interesting implications. If $a_{\rm max}$ is known,
%then the softening may be chosen so as not to dominate the optimally,
%in a manner essentially independent of the number of particles. Of
%course, $a_{\rm max}$ is only well defined if the slope of the inner
%density profile is as shallow (or shallower) than $\beta=1$, which is
%one of the issues that we are trying to address. Nevertheless, this
%result allows us to derive the `converged' region of the mass
%profile {\it a posteriori}, by identifying the radius where the local
%acceleration equals the characteristic value imposed by the softening.
%(ii) Depending on the value of
%$\epsilon$, mass profiles may actually be trustworthy even on scales where
%pairwise interactions are non-Newtonian. Consider for example the case of
%$\epsilon \sim 0.03\, r_{200}$, the largest softening that yields a converged
%result in Figure~\ref{figs:accprof}. In this case the mass profile has converged
%even at distances as small as $r \approx \epsilon$ although fully Newtonian
%pairwise accelerations are only recovered at distances greater than $2\,
%\epsilon$.

To summarize, provided that all other numerical parameters are chosen
appropriately, the effect of the softening on the spherically-averaged mass
profile is to impose a maximum acceleration scale above which results cannot be
trusted. The mass profile of a simulated halo converges at radii where the mean
acceleration does not exceed a characteristic value imposed by the softening,
$a(r)=V_c^2(r)/r\lsim a_{\epsilon}=\chi_{\epsilon} V_{200}^2/{\epsilon}$, where
$\chi_{\epsilon}$ is empirically found to be $\sim 0.5$ if $\epsilon$ is
expressed as a spline-softening scalelength.

\subsection{The Initial Redshift}
\label{ssec:zi}

The starting redshift, $z_i$, determines the overall initial amplitude of
density fluctuations in the simulation box. If $z_i$ is too low, small scales
may already be in the non-linear regime, invalidating the assumptions of the
procedure outlined in \S~\ref{ssec:ics}. Initial redshifts cannot be chosen to
be too high either, since the more uniform the periodic box, the more difficult
the task of evaluating accurate forces in treecodes such as the ones we employ
here becomes. A compromise must therefore be struck between these competing
demands and we derive in this section a simple empirical prescription that
ensures convergence in the mass profiles of simulated CDM halos at $z=0$.

\begin{figure}
\centerline{\psfig{figure=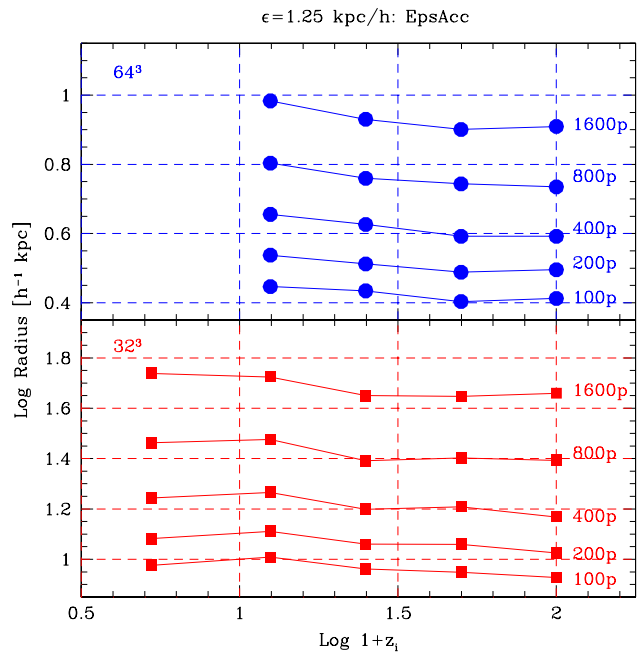,width=250pt,height=250pt}}
\caption{ As Figure~\ref{figs:rvssoft}, but as a function of the
initial redshift of the simulation. Convergence is seen for $z_i\gsim
25$ in the $32^3$ runs and for $z_i \gsim 49$ in the $64^3$
runs. Starting at lower initial redshifts causes halo mass profiles to
develop an artificially low density core. }
\label{figs:rvszi}
\end{figure}

Figure~\ref{figs:rvszi} shows the radii of various mass fractions (at $z=0$) as a function of the initial redshift of the simulation. Top and bottom panels
refer to the same halo, using two different particle numbers in the
high-resolution box: $32^3$ (bottom), and $64^3$ (top).  Each curve is labeled
by the enclosed number of particles. The inner mass profile of the halo
converges as the initial redshift is increased. Convergence to better than
$10\%$ at all radii is achieved for $25< 1+z_i<100$, and even for the highest
$z_i$ tested there is no clear departure from convergence. We have checked
explicitly that this result does not depend on the particular time-stepping
choice; a similar series with the {\tt SgAcc} criterion gives similar results.

\begin{figure}
\centerline{\psfig{figure=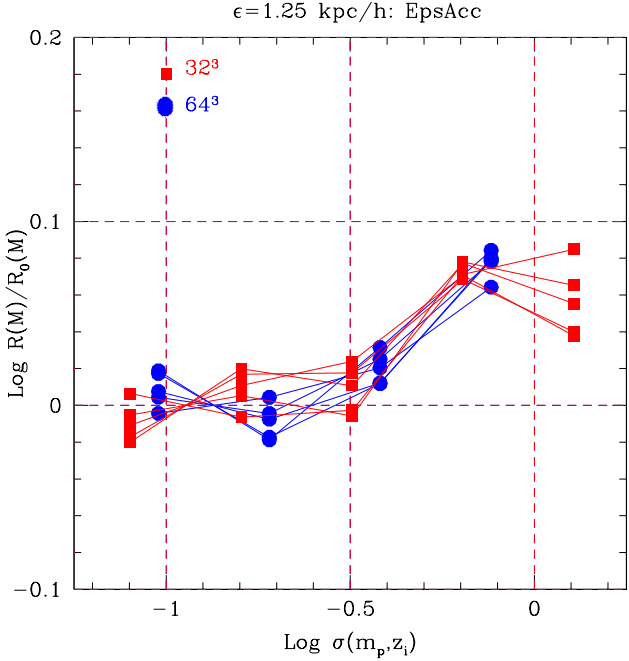,width=250pt,height=250pt}}
\caption{The radii of various mass shells, as in Figure~\ref{figs:rvszi}, but
normalized to the `converged' value of the radius for each shell as a function
of $\sigma(m_p,z_i)$, the linear rms fluctuation on the scale of the particle
mass at $z=z_i$. Note that convergence is achieved at all radii when
$\sigma(m_p,z_i) \lsim 0.3$.}
\label{figs:rvszi_sc}
\end{figure}

The data in Figure~\ref{figs:rvszi} also suggest that convergence may
be achieved at lower $z_i$ when $32^3$ particles are used rather than
$64^3$. A possible explanation for this is presented in
Figure~\ref{figs:rvszi_sc}, where we plot, for each radial shell, the
deviations from the converged value as a function of the (theoretical)
rms mass fluctuation on the smallest resolved mass scale at $z_i$,
$\sigma(m_p,z_i)$ ($m_p$ is the mass of one high-resolution
particle). In terms of this variable, the $32^3$ and $64^3$ results
are indistinguishable, showing convergence down to the $100$-particle
mass shell when $z_i$ is chosen so that $\sigma(m_p,z_i) \lsim
0.3$. This is a simple empirical rule for choosing the starting
redshift that we shall adopt hereafter.

One advantage of this rule is that, for power spectra such as CDM, $\sigma(m_p)$
is only weakly dependent on mass on small scales, so the initial redshift can be
chosen almost independently of the number of particles. For example, even for
the highest number of particles considered in our study ($N_{\rm sbox}=256^3$,
$m_p=6.5\times 10^5 \, h^{-1} M_{\odot}$) the starting redshift condition is
satisfied for $1+z_i \gsim 42$, so that $1+z_i=50$ could be safely used for all
of our simulations, regardless of $N$.

\subsection{Force Accuracy}
\label{ssec:facc}

Accurate forces are an obvious requirement for numerical convergence, and we
investigate here the role of force accuracy parameters in the mass structure of
dark halos. This is important since treecodes are based on approximate multipole
expansion-based methods that are vulnerable to inaccuracies in the force
calculations. Although accuracy can always be improved by adopting, for example,
stricter node-opening criteria, this comes usually at the cost of substantial
loss in computational efficiency. It is therefore important to determine what is
the minimum force accuracy needed to achieve convergence in order to maximize
the efficiency of the simulation.

\begin{figure}
\centerline{\psfig{figure=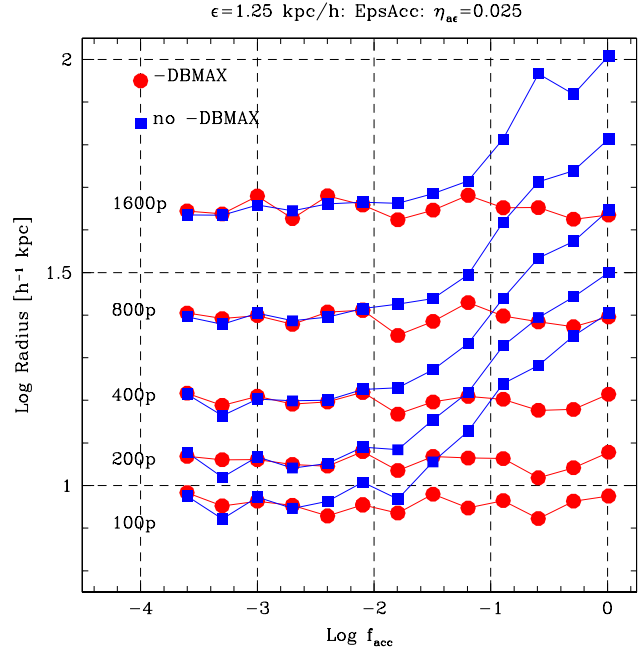,width=250pt,height=250pt}}
\caption{ As Figure~\ref{figs:rvssoft}, but for radii as a function of the 
{\tt GADGET} force accuracy parameter, $f_{\rm acc}$ ({\tt ErrTolForceAcc} in
 {\tt GADGET}'s parameter file). Filled squares show results without enabling 
the extra-accuracy flag {\tt -DBMAX} during compilation. Filled circles show 
results enabling {\tt -DBMAX}. When this flag is on, the effects of $f_{\rm
 acc}$ on the mass are mild, and good convergence is achieved even for rather
 large values of $f_{\rm acc}$. When {\tt -DBMAX} is off, $f_{\rm acc}\lsim 
10^{-3}$ is needed to ensure convergence.}
\label{figs:rfacc}
\end{figure}

Force accuracy is controlled in {\tt GADGET} (in the configuration
used in this study, see \S~\ref{sssec:gadget}) through two main
parameters: a compile-time flag, {\tt -DBMAX}, which, if enabled,
restricts node opening to a list of cells {\it guaranteed} not to
contain the particle under consideration, and by the parameter $f_{\rm
acc}$ (named {\tt ErrTolForceAcc} in {\tt GADGET}'s parameter file),
which controls dynamically the updating of the tree-node opening
criterion (Springel et al. 2001). Figure~\ref{figs:rfacc}
shows the radii of various mass shells in our standard halo as a
function of $f_{\rm acc}$. Filled squares show the results obtained
without setting the {\tt -DBMAX} option in {\tt GADGET}. Convergence
is achieved in this case for quite small values of the accuracy
parameter, $f_{\rm acc} \lsim 0.003$. The reason behind the slow
convergence seen in Figure~\ref{figs:rfacc} appears to be related to
rare but substantial errors incurred in {\tt GADGET}'s tree walking
procedure when the boundaries of open nodes are {\it not guaranteed}
to exclude the particle under consideration (Salmon \& Warren
1993). Disallowing this possibility (i.e., enabling {\tt -DBMAX}
during compilation) leads to much improved convergence relative to the
parameter $f_{\rm acc}$, as can be seen from the filled circles in
Figure~\ref{figs:rfacc}. There is almost no systematic trend with
$f_{\rm acc}$ when {\tt -DBMAX} is enabled, even for $f_{\rm acc}
\approx 1$.

\begin{figure}
\centerline{\psfig{figure=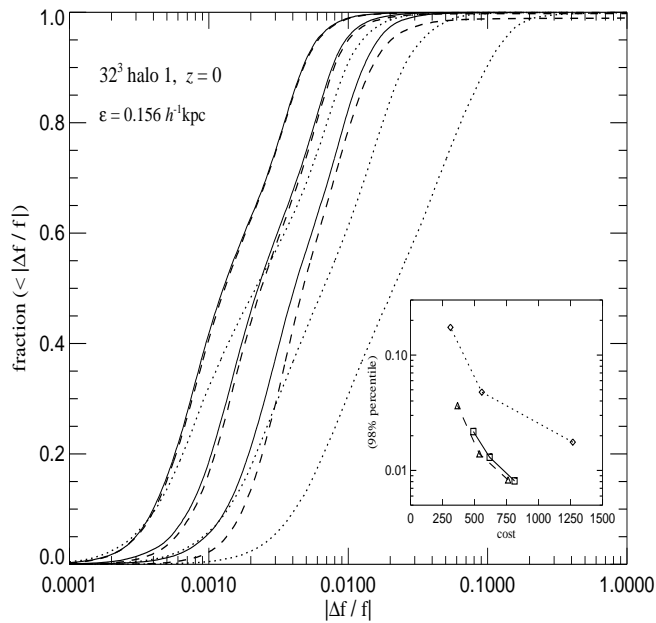,width=250pt,height=250pt}}
\caption{ Cumulative error distributions of {\tt GADGET}'s force
computation for various choices of opening criterion and tolerance
parameter. We used the particle distribution of the z=0 snapshot of a
run with $32^3$ high-resolution particles and measured force errors by
comparing to the result obtained by direct summation. Solid and dashed
lines give the result of opening nodes with the `relative' opening
criterion proposed by Springel et al. (2001), with and without the
{\tt -DBMAX} option (solid and dashed lines, respectively). Results
are shown for tolerance parameters $f_{\rm acc}=0.001$, $0.003$, and
$0.01$ (from left to right). Dotted lines show results for the
traditional BH-opening criterion (dotted lines), with opening angles
$\theta=0.5$, $0.75$, $1.0$ (from left to right). The inset compares
the accuracy obtained for all of these choices as a function of
computational cost. See text for more details.}
\label{figs:facc}
\end{figure}

The main effect of enabling {\tt -DBMAX} is to suppress a tail of large errors
that, although rare, appear to have a significant effect on the final mass
profile. This can be seen in Figure~\ref{figs:facc} where we show the 
cumulative distribution of errors in accelerations computed on a $z=0$ 
snapshot of a simulation with $32^3$ high-resolution particles. Force errors
were measured by comparing with the result obtained by direct summation.  
Solid and dashed lines give the result of opening nodes with the `relative' 
opening criterion proposed by Springel et al. (2001), with and without the 
{\tt -DBMAX} option (solid and dashed lines, respectively). In each case, 
results are shown for tolerance parameters $f_{\rm acc}=0.001$, $0.003$, and
 $0.01$ (from left to right). We also show results for the traditional
 Barnes-Hut opening criterion (dotted lines), with opening angles $\theta=0.5$
, $0.75$, $1.0$ (from left to right).

We have chosen a rather small value of the softening in
Figure~\ref{figs:facc} to emphasize graphically the point that a long
tail of errors may exist when the {\tt -DBMAX} option is not enabled;
note, for example, that errors of up to $100\%$ or larger are present
in this case when $f_{\rm acc}=0.01$ (rightmost dashed line). Such
errors are not present when {\tt -DBMAX} is on (solid lines).  

The inset compares the accuracy obtained for all of these choices as a function
of the invested computational cost. `Accuracy' is here taken as the $98\%$
percentile force error, and the computational cost is measured in terms of the
average number of node-particle interactions per force evaluation. For a given
accuracy, the Barnes-Hut criterion results in higher cost than the criterion
adopted in {\tt GADGET}.

We conclude that enabling {\tt -DBMAX} and adopting $f_{\rm acc} \lsim
0.01$ is sufficient to study the inner structure of dark matter
halos. Alternatively, adopting a redshift-dependent Barnes-Hut node
opening criterion, such as in {\tt PKDGRAV}, where $\theta=0.55$ is
used for $z>2$ and $\theta=0.7$ for $z<2$, seems also to give adequate
results.

\subsection{The Number of Particles}
\label{ssec:npart}

The total number of particles is a critical parameter to choose when running a
cosmological N-body simulation. Since the computation time will scale at best
linearly with $N$, one must try and use as few particles as possible to achieve
the goals of the programme. As mentioned in \S~\ref{sec:intro}, our main goal is
to provide robust and accurate measurements of the circular velocity (or mass)
profile of dark matter halos down to about the inner $1\%$ of the virial
radius. This corresponds to $\sim 2.2 \, h^{-1}$ kpc in the case of the Milky
Way if its halo has the same circular velocity as the disk. This is clearly the
minimum resolution required for meaningful comparison with observed rotation
curves.

In the preceding discussion we have determined the optimal choice of
softening, time stepping, force accuracy, and starting redshift
required to obtain repeatable and robust measurements of the circular
velocity profile of a simulated CDM halo down to radii containing as
few as $100$ particles. Repeatability and robustness relative to these
parameters are, of course, necessary conditions for convergence, but
we must still demonstrate that the results do not depend on the total
number of particles chosen.

How many particles must a region contain so that the circular velocity (or,
equivalently, the mean inner density) converges? 
%However,
%there is conflicting indication in the literature about whether the
%uncertainties in the inner profiles are actually dominated by Poisson
%statistics. For example, Klypin et al. (2001) argue that, provided
%other parameters are chosen wisely, mass profiles can be trusted down
%to the innermost $100$-$200$ particles. Moore et al. (1998, 1999), on
%the other hand, disagree with this conclusion, and present evidence
%that, as the total number of particles increases, the smallest radius
%where convergence is achieved encompasses more and more particles,
%substantially exceeding, in their highest resolution simulations, the
%criterion proposed by Klypin et al. (2001).
We use the lessons from the preceding subsections to explore the
dependence of the mass profile of simulated dark halos on the number
of particles used. We consider only runs which meet the requirements
discussed previously, so that, for each choice of $N$, we shall only
present `converged' results relative to other parameters. Our tests
span an unprecedented range of $512$ in particle number, from $32^3$
to $256^3$ particles in the high-resolution simulation cubes.

\begin{figure*}
\centerline{\psfig{figure=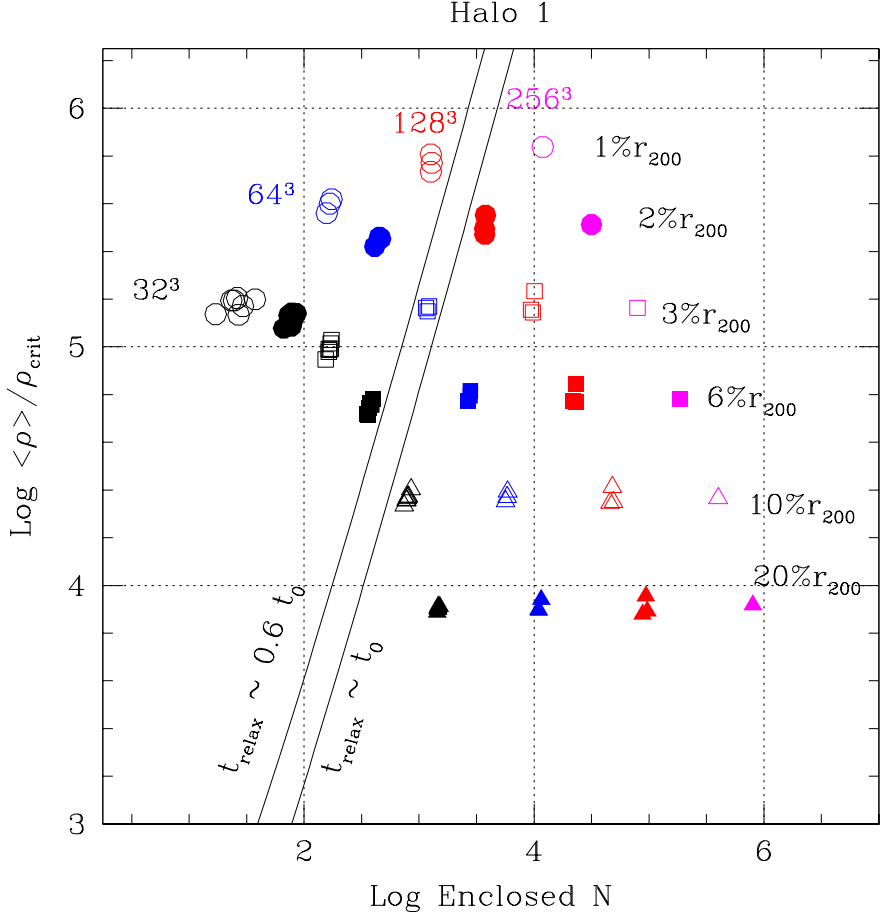,width=350pt,height=350pt}}
\caption{Mean inner density contrast as a function of the enclosed
number of particles in 4 series of simulations varying the number of
particles in the high-resolution box, from $32^3$ to $256^3$. Each
symbol type corresponds to a fixed fraction of the virial radius, as
shown by the labels on the right. The number of particles needed to obtain robust
results increases with density contrast, roughly as prescribed by the
requirement that the collisional relaxation timescale should remain
longer than the age of the universe. According to this, robust
numerical estimates of the mass profile of a halo are only possible to
the right of the curve labeled $t_{\rm relax}\sim 0.6 t_0$.}
\label{figs:rrho_nenc}
\end{figure*}

Our main results are summarized in Figure~\ref{figs:rrho_nenc}, where
we show, as a function of the enclosed number of particles, the mean
inner density contrast measured at various radii from the centre of
the halo. In this figure, for example, solid triangles show the mean
inner density contrast measured at $\sim 20\%$ of the virial radius.
>From left to right, each group of filled triangles indicates the
results of runs with $32^3$, $64^3$, $128^3$, and $256^3$ particles in
the high resolution cube. These runs have $6200$; $49600$; $397000$;
and $3.2\times 10^6$ particles within $r_{200}$, respectively. As the
number of particles increases fewer runs are shown, because of the
increasing computational cost. At the highest resolution, with $256^3$
particles in the high-resolution cube, we have completed only one
simulation. This run is comparable to the highest resolution
simulations reported in the literature so far.

Figure~\ref{figs:rrho_nenc} shows a number of important trends. Consider, for
example, the radius corresponding to $2\%$ of $r_{200}$ (solid circles). In the
$32^3$ runs, this radius contains $1.6\%$ of the halo mass ($\sim 100$
particles). The mass within this radius is seen to increase significantly as the
number of particles increases; the density contrast climbs from $\sim 1.2 \times
10^5$ (in the $32^3$ runs) to $\sim 2.5\times 10^5$ (in the $64^3$ runs) before
stabilizing at $\sim 3 \times 10^5$ when $N_{\rm sbox}$ reaches $128^3$ and
$256^3$. Clearly, $100$ particles are {\it not enough} to trace reliably the
mass profile of a simulated halo, in disagreement with the conclusions of Klypin
et al. (2001), who argue that $100$-$200$ particles suffice to resolve the inner
mass profile when other parameters are chosen properly.

The situation is different for the $1000$-particle radius in the
$32^3$ runs, which correspond to about $10\%$ of the virial
radius. The density contrast within this radius is $\sim 2.5\times
10^4$, and remains essentially unchanged as the number of particles
increases by a factor of $512$. The data presented in
Figure~\ref{figs:rrho_nenc} thus support the conclusions of Moore et
al (1998): resolving regions closer to the centre, where the
density contrast is higher, demands increasingly large particle
numbers. Although $300$ particles in the $32^3$ runs are almost enough
to resolve the $6\%$ radius, they fall well short of what is needed to
resolve the much higher overdensities characteristic of the $1\%$
radius.

How many particles are needed to resolve a given radius? Moore et al. (1998)
propose that converged regions are delineated by (one-half) the mean
inter-particle separation within the virial radius, $0.5 \, (4\pi/3N_{200})^{1/3}
r_{200}$, whereas Fukushige \& Makino (2001)  suggest that the innermost
resolved radius cannot be smaller than the radius where the two-body relaxation
time becomes shorter than the age of the universe.

Our results appear to favor the latter interpretation. For example,
the criterion of Moore et al. would predict that the $32^3$ runs could
be trusted down to $4.5\%$ of the virial radius, but it is clear from
Figure~\ref{figs:rrho_nenc} that convergence in this case is achieved
only for radii beyond $6\%$ of $r_{200}$. On the other hand, all
simulations can be seen to converge at radii larger than the radius
where the average collisional relaxation time roughly matches the age
of the universe. This is shown by the (almost vertical) line labeled
$t_{\rm relax} \sim t_0$, where we define
\begin{equation} 
\label{eq:trelax}
{t_{\rm relax}(r) \over t_{\rm circ}(r_{200})} =
{N\over 8\ln{N}} {r/V_c \over r_{200}/V_{200}}  =
{\sqrt{200} \over 8} {N\over \ln{N}} \left({{\bar \rho} \over \rho_{\rm crit}}\right)^{-1/2},
%{\pi \sqrt{2}\over 8\ln(\Lambda_{\rm C})} N(r)\, t_{dyn}(r).
%
\end{equation}
$t_{\rm circ}(r_{200})\sim t_0$, and $N=N(r)$ is the enclosed number
of particles. For reference, the curve on the left indicates $t_{\rm
relax}=0.6 \, t_{\rm circ}(r_{200}) \sim 0.6 \, t_0$.  As shown in
Figure~\ref{figs:rrho_nenc}, the density profile converges at radii
that enclose enough particles so that $t_{\rm relax}(r) \gsim 0.6 \,
t_0$.

We emphasize that this criterion is mainly empirical, and does not necessarily
imply that particles in regions where the relaxation time is shorter than $\sim 0.6 \, t_0$ actually evacuate the central regions as a result of two-body
encounters. Indeed, one would expect the inner mass profile to evolve as a
result of collisions on the much longer `evaporation' timescale, $t_{\rm
evap}\approx 136 \, t_{\rm relax}$ (Binney \& Tremaine 1987), a proposition that
finds support in simulations of the evolution of isolated equilibrium $N$-body
systems (Hayashi et al 2002). In addition, the heating rate near the
centre is likely dominated by the presence of substructure rather than by
particle-particle collisions, complicating the interpretation. Our result is
thus reminiscent of the work of Weinberg (1998), who emphasizes the difficulty
of achieving the collisionless limit in $N$-body systems and the possibility
that fluctuation noise may lead to relaxation effects important on all scales.

Despite this difficulty, it seems clear from Figure~\ref{figs:rrho_nenc} that
resolving density contrasts exceeding $10^6$ requires $\gsim 3000$ particles
within that radius, or over $3$ million particles within the virial
radius. Providing robust numerical predictions of the mass structure of cold
dark matter halos on scales that can be compared directly with observations of
individual galaxies is thus a very onerous computational task.

\subsection{Optimal parameters - a worked example}
\label{ssec:workex}

The many considerations discussed in the previous sections make the selection of
optimal parameters for any given N-body run a delicate and complicated
business. It may be helpful to go through how one might choose optimal
parameters for a specific calculation, for example a simulation like the largest
one ($N_{\rm sbox}=256^3$) we consider in this paper. This run has $\sim 3\times
10^6$ particles within the virial radius at $z=0$, and is the largest we can
easily carry out with resources currently available to
us. Figure~\ref{figs:rrho_nenc} and the discussion in \S~\ref{ssec:npart}
suggest that this number of particles should be sufficient to get converged
results down to about $r_{conv} =0.005 \, r_{200}$. Equation~\ref{eq:epsopt}
suggests that a softening parameter $\epsilon = 0.0025 \, r_{200}$ will be near
optimal for getting an efficient integration almost unaffected by discreteness
effects. As Figure~\ref{figs:accprof} demonstrates, this softening is small
enough relative to our target $r_{conv}$ that it should not compromise the
radial structure. For these parameters, equation~\ref{eq:tcrconv} and
Figure~\ref{figs:tcprof} then show that a single-timestep integrator should be
able to converge in about $5000$ equal steps, although we note that this depends
on the detailed inner structure of the halo, which is what we are trying to
measure. In practice, a series of runs where the number of particles is
gradually increased, is desirable to fine-tune the choice of
timestep. Alternatively, the discussion of
\S~\ref{sec:multistep} implies that for our preferred multi-timestep integrator
({\tt EpsAcc}) $\eta = 0.15$ should be small enough to ensure convergence. The
discussion of \S~\ref{ssec:zi} shows that it should be safe to start the
integration at $z_i=49$.

\section{The  Circular Velocity Profile of a $\Lambda$CDM Halo}
\label{sec:vprof}

\begin{figure}
\centerline{\psfig{figure=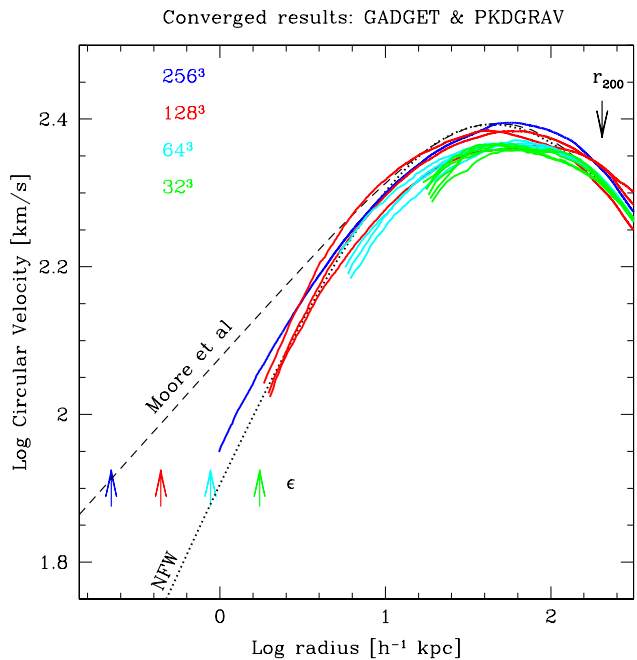,width=250pt,height=250pt}}
\caption{Circular velocity profiles of `converged' runs with different
number of high-resolution particles. Profiles are only plotted for radii where
the convergence criteria derived in this paper are satisfied. Several curves are
shown for the cases of $32^3$, $64^3$, and $128^3$ particles, corresponding to
runs where all other numerical parameters take converged values. For clarity, a
small selection of runs have been chosen; those with softenings indicated by the
small vertical arrows. The convergent profile that emerges for this halo is
roughly independent of the number of particles and resembles closely the model
proposed by NFW, with $c=10$. For this halo, steeply-cusped density profiles are
disfavoured. The profiles labeled `Moore et al' and `NFW' have been matched at
the peak of the circular velocity profile.}
\label{figs:vprof}
\label{lastpage}
\end{figure}

Finally, we use the convergence lessons derived above to analyze
briefly the inner circular velocity profile of the $\Lambda$CDM halo
considered here. The results of `converged' runs are shown in
Figure~\ref{figs:vprof}. Each profile is shown only for radii
considered converged according to the criteria discussed
above. Plotted this way, all profiles, independent of the number of
particles, seem to agree to within $\sim 10\%$ at all radii. The
circular velocity increases from the virial radius inwards, reaches a
maximum and then drops gradually towards the centre, following closely
the dotted line that represents an NFW profile with concentration
$c=10$. This value of the concentration agrees reasonably well with
the results of NFW and of Eke, Navarro \& Steinmetz (2001), who find
$c\approx 8$-$9$ for a halo of this mass.  Near the centre, the
profile is seen to deviate significantly from the steeply cusped
profile approaching a central slope of $\beta=1.5$ proposed by Moore
et al. (1998), and agrees better with shallower central slopes such as
that of the NFW model.

We emphasize that there is little evidence for convergence to a power-law
density profile near the centre, and that the profile keeps getting shallower
down to the innermost point that our procedure deems converged. Can our results
be used to place meaningful constraints on the asymptotic inner slope?  At
$r_{\rm min}\sim 1\, h^{-1}$ kpc, the smallest radius resolved in our
highest-resolution run ($N_{\rm sbox}=256^3$), both the local and cumulative
density profiles are robustly determined{\footnote{Convergence in the local
density actually extends to radii smaller than the minimum converged radius for
the more stringent cumulative density.}}: $\rho(r_{\rm min})/\rho_{\rm
crit}=9.4\times 10^5$, and ${\bar \rho}(r_{\rm min})/\rho_{\rm crit}\approx
1.6\times 10^6$. These values can be combined with the requirement of mass
conservation to place an upper limit to the inner asymptotic slope of the
density profile, $\beta < 3 (1-\rho(r_{\rm min})/{\bar \rho(r_{\rm
min})})=1.2$. In other words, there is not enough mass within $r_{\rm min}$ to
support a power-law density profile with slope steeper than $\beta=1.2$. We note
that this conclusion depends sensitively on our ability to resolve the innermost
$1\, h^{-1}$ kpc. If $r_{\rm min}$ were just two or three times larger the same
exercise would not be able to rule out slopes as steep as $\beta=1.5$.

In summary, our results argue strongly against the very steep central
cusps advocated by Moore et al. (1998, 1999), Ghigna et al. (1998, 2000)
and Fukushige \& Makino (1997, 2001). We are in the process of
augmenting our sample of halos in order to firm up this conclusion, so
will defer a detailed analysis of this issue to a later paper in this
series.

\section{Conclusions}
\label{sec:conclusions}
We have performed a comprehensive series of convergence tests designed
to study the effect of numerical parameters on the structure of
simulated cold dark matter halos. Our tests explore the influence of
the gravitational softening, the time-stepping algorithm, the starting
redshift, the accuracy of force computations, and the number of
particles on the spherically-averaged mass profile of a galaxy-sized
halo in the $\Lambda$CDM cosmogony. We derive, for each of these
parameters, empirical rules that optimize their choice or, when those
choices are dictated by computational limitations, we offer simple
prescriptions to assess the effective convergence of the mass profile
of a simulated halo. Our main results can be summarized as follows:

\begin{enumerate}

\item 
{\it Timestep and Discreteness Effects.}  The number of timesteps required to
achieve convergence depends primarily on the orbital timescale of the region to
be resolved, but may also be sensitive to the number of particles and the
gravitational softening, unless these parameters are chosen so that discreteness
effects are unimportant. This requires the gravitational softening to be large
enough so that the maximum acceleration during two-body encounters does not
exceed the minimum mean field acceleration in the halo, $\epsilon \gsim
\epsilon_{\rm acc}=r_{200}/\sqrt{N_{200}}$. Empirically, we find that
$\epsilon\approx \epsilon_{\rm opt}=4 \, \epsilon_{\rm acc}$ gives good
results. When this condition is satisfied, the minimum converged radius, $r_{\rm
conv}$, is given by the condition that the circular orbit timescale should be
long compared to the timestep, $t_{\rm circ}(r_{\rm conv}) \approx 15
\left({\Delta t/ t_0}\right)^{5/6} t_{\rm circ}(r_{200})$.  {\it Substantially
smaller} timesteps are needed if $\epsilon<\epsilon_{\rm opt}$. Dark matter
densities at $r< r_{\rm conv}$ may be under- or over-estimated, depending on the
integrator and timestepping schemes used. For example, constant-timestep {\tt
GADGET} runs develop artificially dense, `cuspy' cores in poorly resolved
regions, indicating that the approach to convergence is not always
monotonic. This emphasizes the importance of comprehensive convergence tests
such as the ones presented here to validate the results of numerical studies of
the inner structure of CDM halos.

\item
{\it Fixed Timestep versus Adaptive Multi-Stepping.}  Of the several adaptive,
multiple time-stepping criteria that we considered, we have found best results
when timesteps are chosen to depend explicitly on the gravitational softening
and on the acceleration, $\Delta
t_i=\eta_{a\hskip-0.5pt\epsilon}\sqrt{\epsilon_i/a_i}$, with
$\eta_{a\hskip-0.5pt\epsilon} \sim 0.2$. Experiments with time-stepping choices
that do not include explicitly the gravitational softening require the value of
the corresponding $\eta$ to be reduced as $\epsilon$ is reduced below the
optimal value in order to obtain convergence. In terms of computational cost, we
find that multi-time-stepping criteria significantly outperform the use of a
single timestep for all particles only for softenings well below the optimal
value.

\item 
{\it Gravitational Softening.} The choice of gravitational softening is found to
impose a maximum acceleration scale above which simulation results cannot be
trusted. This acceleration scale appears to depend mainly on the circular
velocity of the halo and on the gravitational softening scale, and is given by
$a_{\epsilon}=\chi_{\epsilon}\, V_{200}^2/\epsilon$, with $\chi_{\epsilon} \sim
0.5$. For {\it given particle number}, convergence to better than $10\%$ in the
mass profile is obtained at radii greater than $\epsilon$ that also contain more
than $100$ particles and where the acceleration criterion is satisfied:
$a(r)=V_c(r)^2/r \lsim a_{\epsilon}$.

\item 
{\it Starting Redshift.} The mass profiles of simulated dark halos converge
provided that the initial redshift is chosen so that the theoretical (linear)
rms fluctuations on the smallest resolved mass scale, $m_p$ (the mass of one
high-resolution particle) is $\sigma(m_p,z_i) \lsim 0.3$. Since $\sigma(m_p)$ is
a weak function of mass on subgalactic mass scales for CDM-like power spectra,
this criterion indicates that a modest starting redshift, such as $1+z_i\approx
50$ is appropriate for particle masses as low as $m_p \sim 10^5 \, h^{-1}
M_{\odot}$ in the $\Lambda$CDM cosmogony.

\item
{\it Force Accuracy.} The mass profiles of simulated CDM halos are quite
sensitive to the accuracy of the force calculations, and convergence requires
care in the choice of node opening criteria in the treecodes used in our
study. Poor force accuracy leads to the development of artificially low density
cores.  In the case of {\tt GADGET}, for example, we find that even occasional
large errors in the forces may lead to noticeable deviations from converged
profiles. To avoid this, it is necessary to choose tree-walking parameters that
curtail drastically the tail of the most deviant force calculations, however
rare. In {\tt GADGET} this can be achieved by activating the compiler option
{\tt -DBMAX}. Using up to hexadecapole terms in the node potential expansion and
setting a redshift dependent tree-node opening criterion, as in {\tt PKDGRAV},
where $\theta=0.55$ is chosen for $z>2$ and $\theta=0.7$ for $z<2$, seems also
to work well.

\item 
{\it Particle Number.}  In order to achieve convergence in the mass profile,
enough particles must be enclosed so that the average two-body relaxation
timescale within the region is comparable or longer than the age of the
universe. We find empirically that the condition, $t_{\rm relax}(r)\gsim 0.6
\, t_0$, describes converged regions well. Since $t_{\rm relax}$ is roughly
proportional to the enclosed number of particles times the local dynamical
timescale, resolving regions near the centre, where density contrasts are high
and dynamical timescales are short, requires substantially more particles than
resolving regions more distant from the centre. Of order $3000$ enclosed
particles are needed to resolve regions where the density contrast reaches
$10^6$. On the other hand, density contrasts of order $10^{4.5}$ require only
$100$ enclosed particles for numerical convergence. Resolving radii of order
$0.5\%$ of the virial radius in the first case requires of order $3 \times 10^6$
particles within the virial radius.

\end{enumerate}

For most simulations, the most stringent convergence criterion is the relaxation
timescale condition on the number of particles. This implies that there is
little choice but to strive for the largest possible $N$ when studying the inner
regions of dark matter halos. This limit is dictated by the available computer
resources. Choosing the optimal softening for the adopted number of particles
then minimizes the number of timesteps needed to achieve convergence down to the
radius where $t_{\rm relax}(r)\gsim 0.6 \, t_0$. The precise number of timesteps
cannot be determined ahead of time, since $t_{\rm relax}(r)$ depends on the
detailed structure of the halo, which is what we are trying to measure. This
implies that a series of simulations where the number of particles is increased
gradually is advisable in order to ensure that optimal parameters are chosen for
the highest-resolution run intended.

We have applied our convergence criteria to a $\sim 205$ km s$^{-1}$
$\Lambda$CDM halo in order to investigate the behaviour of the inner slope of
the density profile. We find that the slope of the spherically-averaged density
profile, $\beta=-{\rm d}\log({\rho})/{\rm d}\log(r)$, becomes increasingly
shallow inwards, with little sign of approach to an asymptotic value.  At the
smallest radius that we consider resolved in our highest-resolution ($256^3$)
simulation ($r_{\rm min} \sim 1 \, h^{-1}$ kpc $\approx 0.005\, r_{200}$), the
local and cumulative density contrasts are robustly determined, $\rho(r_{\rm
min})/\rho_{\rm crit}=9.4\times 10^5$, and ${\bar \rho}(r_{\rm min})/\rho_{\rm
crit}\approx 1.6\times 10^6$. These values can be combined with the requirement
of mass conservation to place an upper limit to the inner asymptotic slope of
the density profile, $\beta < 3 \, (1-\rho(r_{\rm min})/{\bar \rho(r_{\rm
min})})=1.2$, although it is possible that the slope may actually become even
shallower near the centre, as suggested recently by Taylor \& Navarro (2001).

Our results thus argue against the very steep values for the asymptotic central
slope ($\beta\approx 1.5$) claimed recently by Moore et al. (1998, 1999), Ghigna
et al. (1998, 2000), and Fukushige and Makino (1997, 2001). The reasons for this
disagreement are unclear at this point, since there are substantial differences
in the halo mass, numerical techniques, and cosmological model adopted, which
hinder a direct comparison between our results and theirs. For example, the work
of Moore et al. (1998) and Ghigna et al. (2000) differs from ours in mass scale
(they simulated a galaxy cluster while we target a galaxy-sized halo) and in
cosmology (they adopted an Einstein-de Sitter CDM cosmogony, whereas we adopt
the $\Lambda$CDM model). 
%The models of Fukushige \& Makino (2001), on the other
%hand, use initial conditions that are not drawn from a self-consistent
%cosmological context.

Finally, the difference between the conclusions from various authors may just
reflect the fact that each group applies different criteria to the
identification of the regions deemed trustworthy. We note that models with the
very steep ($\beta\sim 1.5$) inner slopes proposed by the Moore et al group and
with the shallower slopes that we find here are almost indistinguishable if we
restrict our analysis to radii $\gsim 2\%$ of the virial radius. Probing radii
within the inner $1\%$ of the virial radius seems required to shed light on this
controversy. Further simulation work with resolution adequate to address this
issue in detail is currently underway.

\section*{Acknowledgements}
The Natural Sciences $\&$ Engineering Research Council of Canada (NSERC) and the
Canadian Foundation for Innovation have supported this research through various
grants to JFN. This work has been supported by the UK PPARC and by the EC
through a European Research Network.  Special thanks to Colin Leavitt-Brown for
expert assistance with the IBM-SP3 supercomputing facilities at the University
of Victoria. Many of the simulations were carried out on facilities
at the Edinburgh Parallel Computing Centre and the Rechenzentrum, Garching

\begin{appendix}
\medskip
\section{The Generation of Cosmological Initial Conditions}

Periodic boundary conditions are usually adopted in cosmological simulations for
reasons of convenience.  The assumption of periodicity implies that the
simulation volume as a whole has to have precisely the mean density, a
requirement that places restrictions on the size of the region and on the
redshifts at which a particular simulation may be considered reliable. On the
other hand, with periodic boundaries the density field can be expanded as a sum
over a discrete set of periodic plane waves.  For a simulation volume which is
cubic, the Fourier transform of the density field has the form of a cubic grid
in Fourier space.  The discrete nature of the power spectrum thus makes it easy
to set up Gaussian density fields.

The aim of our initial conditions generating procedure is to provide a particle
realization of a Gaussian density field with the chosen power spectrum, $P(k)$,
on scales and at redshifts where linear theory is applicable. Our procedure
follows closely that described in Efstathiou et al. (1985), where further details
may be found. As in Efstathiou et al. (1985), we use the Zel'dovich approximation
to perturb particles from a uniform cubic grid,
\begin{equation}
\label{eq:xqt}
{\bf x}(t)={\bf q} - b(t) {\bf \psi}({\bf q}), 
\end{equation}
where ${\bf x}$ is the comoving Eulerian coordinate of the particle,
${\bf q}$ is the Lagrangian coordinate denoting the particle's
unperturbed position in the grid, $b(t)$ is the linear growth factor,
and ${\bf \psi}$ is a function that describes the spatial structure of
the density field. The function ${\bf \psi}$ can be expressed in terms
of the acceleration field at time $t$,
\begin{equation}
\label{eq:psi}
{\psi}({\bf q})=- {{\bf F}({\bf q},t) \over ma^2(a{\ddot b}+2{\dot b} {\dot a})},
\end{equation}
where ${\bf F}$ is the force field, $a(t)$ is the expansion factor,
$m$ is the particle mass, and a dot denotes a time derivative.

In practice, a realization of the desired fluctuation distribution is created in
Fourier space, with random phases and normally distributed amplitudes for the
real and imaginary components of each mode. We then multiply by an appropriate
Green's function, and transform back to obtain the potential on a spatial
mesh. This potential is differenced to obtain ${\bf F}({\bf q},t)$ which,
together with eq.~\ref{eq:xqt} and eq.~\ref{eq:psi}, gives the displacement
field required to generate the desired density fluctuations from a uniform
distribution.

Once the displacements from the unperturbed positions have been computed,
velocities are assigned to the particles assuming that only growing modes are
present. The peculiar velocity is then simply proportional to the displacement
vector,
\begin{equation}
\label{eq:xdot}
{\dot {\bf x}}=-{\dot b}\, {\bf \psi}({\bf q})
\end{equation}
In cases such as CDM, where there is significant power on all scales, it is
important to avoid unrealistically large initial velocities that may result from
large amplitude fluctuations on small scales. Thus we assign peculiar velocities
only after recalculating the accelerations using the perturbed particle
positions and using eq.~\ref{eq:psi} to re-estimate ${\bf
\psi}({\bf q})$.

We use a cubic grid distribution of particles to represent a uniform density
distribution for all simulations reported here, but it is also possible to use a
`glass' for the unperturbed configuration.  As discussed by White (1994), this
is a better choice for highly aspherical perturbations, and avoids artifacts
that arise from the existence of `preferred' (Cartesian) directions in cubic
grids. This is especially important when attempting to simulate very low mass
halos in CDM cosmogonies, since on those scales the mass fluctuation spectrum is
nearly `flat' ($P(k)$ approaches $k^{-3}$) and collapse proceeds almost
simultaneously on many different mass scales in a network of sheets and
filamentary structures.

\section{Mass Refinement Technique}

As discussed above, simulations of periodic boxes are only reliable provided
that the box is large enough so that perturbations on scales comparable to the
box size are still linear by the present time. This sets a minimum size for
periodic boxes designed to be run to $z=0$ in a $\Lambda$CDM universe. For
example, in the case considered in this paper (see \S~\ref{ssec:ics}) the size
of the periodic box is $L_{\rm box}=32.5 \, h^{-1}$ Mpc ($M_{\rm box}=9.533
\times 10^{15} \, \Omega_0 \, h^{-1} \, M_{\odot}$), and the variance at $z=0$ is
already $\sigma(M_{\rm box}) \approx 0.3$, at the limit of what may be used to
obtain a good representation of large scale structure in this cosmogony. Clearly
boxes smaller than $32.5 \, h^{-1}$ Mpc cannot capture the correct statistical
properties of the dark matter distribution at $z=0$. Our original low-resolution
simulation was carried out with $128^3$ particles.

Even if $512^3$ particles were used in such a box (at the limit of what is
possible with today's largest supercomputer if many thousands of timesteps are
needed) the mass per particle in a $32.5 \, h^{-1}$ Mpc box would be $m_p=7.1
\times 10^7 \, \Omega_0 \, h^{-1} M_{\odot}$, and a galaxy-sized, $10^{12} \,
h^{-1} M_{\odot}$ halo would only contain slightly more than $10,000$
particles. A dwarf galaxy halo would have fewer than $1,000$ particles. Clearly,
a different technique is required in order to improve the mass and spatial
resolution of the calculation while at the same time accounting properly for the
effects of large scale structure.

The technique most widely adopted so far selects a few systems identified from
the final configuration of the periodic box and resimulates the whole box, with
coarser resolution everywhere except in the selected regions. This technique has
been used in a number of cosmological simulations (see, e.g., Katz \& White
1993, Navarro \& White 1994, Evrard, Summers \& Davis 1994, Moore et al. 1998),
and has become common in high-resolution simulation work targeted at individual
systems. The price one pays with this procedure is that to build a statistically
significant sample of halos entails running many different simulations and there
is always the possibility of introducing biases during the selection procedure.
Having identified a halo in the periodic box for resimulation, all particles
within $\sim 2\, r_{200}$ from its centre are traced back to the initial
conditions and their positions on the original cubic grid are recorded. A box of
size $L_{\rm sbox}$ enclosing all of these particles is then defined. 

A displacement field is generated for $N_{\rm sbox}=256^3$ particles in this new
box using a two-step procedure that allows for inclusion of fluctuations from
the original periodic box.  In the first step, displacements for the $N_{\rm
sbox}$ particles are calculated using the {\it same} Fourier representation as
in the original box, except for the contribution from wavelengths shorter than a
characteristic scale, $d_{\rm lcut}$.  Typically, $d_{\rm lcut}$ is chosen to be
the shortest wavelength in the original box, $d_{\rm lcut} = 2L_{\rm box}/N_{\rm
box}^{1/3} = 2
\times 32.5/128 \, h^{-1} {\rm Mpc} \sim 0.5 \, h^{-1}$ Mpc, which is
the Nyquist wavelength of the low-resolution particle grid.  We truncate the
waves at a boundary which is cubical in Fourier space.
%The only reason for preferring a
%larger value of $d_{\rm lcut}$ is that care must be taken to interpolate the
%long waves sufficiently smoothly - otherwise one can introduce artificial
%distortions to the power spectrum.  Having a higher value of $d_{\rm lcut}$ than
%the minimum means that the field to be interpolated is intrinsically smoother --
%though using too large a value carries the danger than the resimulation will not
%reproduce the original structure at all.

It is important to ensure that the displacements due to the long wavelength
Fourier components are applied to the high resolution particles in a
sufficiently smooth fashion to avoid introducing significant spurious power.
Computing the displacements by simple finite differencing of the potential, as
is the case for the large periodic simulation box, is inadequate in this context
unless an impractically large mesh is deployed. A better way is to compute the
individual components of the displacement field one at time, using the
appropriate Green's functions, and to interpolate the displacement components
themselves, by trilinear interpolation to the individual particle positions.
The use of trilinear interpolation ensures that the displacement field is
continuous--which in itself is sufficient to avoid spurious non-linear features
being introduced.  The larger the mesh used the more accurate is the
interpolation.  For the simulations reported here a $512^3$ mesh was used and
proved satisfactory.

In the second step, fluctuations are generated on scales smaller than $d_{\rm
lcut}$, down to the Nyquist frequency of the high-resolution box.  The new
displacement field is periodic within $L_{\rm sbox}$, and can be vector added to
the large-box displacements in order to obtain final perturbed positions for all
particles within the high-resolution box.  Trilinear interpolation is once again
used to assign the short wave components of the displacement field to the
particles. Peculiar velocities proportional to the displacements are then
assigned using the Zel'dovich approximation and assuming that only growing modes
are present.

Following this procedure, a realization of the displacement field of $256^3$
particles is created and stored for each halo. Finally, the high-resolution box
is inserted in the large periodic box after removal of all overlapping particles
%
%{\footnote{ Total mass is conserved by choosing sizes for the small box that are
%multiple integers of $L_{box}/N_{box}^{1/3}$}}.
%
Not all particles in the small box will end up near the system of interest, so
the location on the original grid of selected particles is used to identify an
`amoeba-shaped' region within the cube that is retained at full
resolution. Regions exterior to the `amoeba' are coarse-sampled using particle
masses which increase with distance from the region of interest
(Figure~\ref{figs:icbox}). The sampling is typically done by binning together
cubes of $2^{3n}$ neighboring particles from the initial grid (where $n$ is an
integer). This allows us to concentrate numerical resources within our selected
object without compromising the contribution from larger scales to the tidal
field acting on the system.  Because of the non-spherical nature of the collapse
of dark halos, accurate simulation of the formation of a single system incurs a
significant overhead. Even after all this optimization, at most $1$ in $3$
particles in the amoeba region ends up within the virial radius of the system
considered.

The success of our procedure may be gauged by computing the power spectrum from
the displaced particle positions and comparing it with the theoretical power
spectrum that we are trying to generate. Figure 1 shows the desired theoretical
power spectrum, the power spectrum measured from the parent simulation, and the
power spectrum measured from a high-resolution box created in the manner
outlined above.  The power spectra are shown at $z=49$.  In this case, the high
resolution box is $5.08 \, h^{-1}$ Mpc on a side and is sampled with $256^3$
particles, with individual masses of $6.5 \times 10^5 \, h^{-1} M_{\odot}$. The
power spectrum of the small box is actually determined for a cube of $4.3 \,
h^{-1}$ Mpc excised from the middle of the high resolution region.  The excised
region would contain $216^3$ particles if it had precisely mean density. The
density field is assigned to a $432^3$ mesh using a cloud-in-cell assignment
scheme and periodic boundary conditions forced.  Forcing periodicity does not
significantly distort the power spectrum for modes small compared to the
fundamental mode of the cube.  The power spectrum is then computed from the
Fourier transformed density field.  The power from individual modes is binned in
shells of constant cubical wavenumber ($k_{cubical}={\rm
max}(|k_x|,|k_y|,|k_z|)$). Plotting the power spectrum using the cubical
wavenumber highlights discrepancies more sharply than the more usual spherical
binning. The good agreement between the theoretical power spectrum and that
measured in our realizations gives us confidence that our simulations faithfully
follow the formation of a dark matter halo in the $\Lambda$CDM cosmogony.

\end{appendix}

\end{document}